\documentclass[12pt,letterpaper]{article} 
 
\global\def\draftcontrol{0} 
 
   \def\versionno{Some  properties of a confining field  
theory at finite temperature via gauge/gravity duality } 
\catcode`\@=11 
 
\expandafter\ifx\csname draftcontrol\endcsname\relax\global\def\draftcontrol{0} 
\fi 
 
{\count255=\time\divide\count255 by 60 
\xdef\hourmin{\number\count255} 
\multiply\count255 by-60\advance\count255 by\time 
\xdef\hourmin{\hourmin:\ifnum\count255<10 0\fi\the\count255}} 
\def\draftdate{\number\month/\number\day/\number\year\ \ \ \hourmin } 
 
\newcommand\makepapertitle{\par 
  \begingroup 
    \renewcommand\thefootnote{\@fnsymbol\c@footnote}%
    \def\@makefnmark{\rlap{\@textsuperscript{\normalfont\@thefnmark}}}%
    \long\def\@makefntext##1{\parindent 1em\noindent 
            \hb@xt@1.8em{%
                \hss\@textsuperscript{\normalfont\@thefnmark}}##1}%
     \newpage 
     \global\@topnum\z@   
     \@makepapertitle 
     \thispagestyle{empty}\@thanks 
  \endgroup 
  \setcounter{footnote}{0}%
  \global\let\thanks\relax 
  \global\let\makepapertitle\relax 
  \global\let\@makepapertitle\relax 
  \global\let\@thanks\@empty 
  \global\let\@author\@empty 
  \global\let\@date\@empty 
  \global\let\@title\@empty 
  \global\let\title\relax 
  \global\let\author\relax 
  \global\let\date\relax 
  \global\let\and\relax 
  \def\version{\let\version\@version\@gobble} 
} 
\def\@makepapertitle{%
  \newpage 
   \ifnum\draftcontrol=1 {} 
   \version\versionno 
   \vskip 3em%
   \else 
   \hfill\hbox to 3cm {\parbox{4cm}{\@pubnum}\hss}%
   \vskip 3em%
   \fi 
   \begin{center}%
   \let \footnote \thanks 
     {\LARGE {\@title}}%
     \vskip 1.5em%
     {\normalsize
       \lineskip .5em%
       \begin{tabular}[t]{c}%
         \@author 
       \end{tabular}\par}%
     \vskip 1.5em%
     {\@bstract}%
     \end{center}%
     \vskip 1.5em  
     \@date%
   \par 
} 
 
\gdef\@pubnum{} 
\def\pubnum#1{%
  \gdef\@pubnum{#1}} 
 
\gdef\@bstract{} 
\def\Abstract#1{%
  \gdef\@bstract{%
   \parbox{\textwidth-0pc}{%
   \centerline{\bf Abstract}\penalty1000%
\noindent
\renewcommand\baselinestretch{1.0}%
{#1}}} 
}

\def\ps@paper{\let\@mkboth\@gobbletwo%
     \ifnum\draftcontrol=1 
        \def\@oddfoot{\hbox to \textwidth{\tiny \versionno \hfil\tiny\draftdate}%
        \hskip -\textwidth \hbox to \textwidth{\hfil\rm\thepage\hfil}}%
     \else\def\@oddfoot{\hbox to \textwidth{\hfil\rm\thepage\hfil}} 
     \fi 
     \let\@evenfoot\@oddfoot 
} 
 
\def\@version#1{\ifnum\draftcontrol=1 
\typeout{}\typeout{#1}\typeout{} 
\vskip3mm\centerline{\hbox{\fbox{\normalsize{\tt DRAFT -- #1 -- } 
                   {\draftdate}}}}\vskip3mm 
\fi} 
\let\version\@version 
\long\def\eqlabel#1{\ifnum\draftcontrol=1 
                    \tag@false  
                    \tag*{(\theequation) \hbox to -0.2cm{\hspace{0cm}\small{#1}\hss}} 
                    \refstepcounter{equation}  
                    \edef\@currentlabel{\theequation} 
                    \ltx@label{#1}          
                    \else 
                    \label{#1} 
                    \fi 
                    } 
\let\st@bibitem\@bibitem 
\let\st@lbibitem\@lbibitem 
\ifnum\draftcontrol=1 
  \def\@bibitem#1{%
    \st@bibitem{#1}\a@@label{#1}\ignorespaces} 
  \def\@lbibitem[#1]#2{%
    \st@lbibitem[#1]{#2}\a@@label{#2}\ignorespaces} 
  \def\a@@label#1{%
    \gdef\a@lab{\smash{\normalfont\small#1}} 
    \ifvmode 
      \if@inlabel 
        \global\setbox\@labels\hbox{%
          \llap{\a@lab\let\a@lab\relax 
                \kern\@totalleftmargin\kern\marginparsep}%
          \box\@labels}%
      \fi 
    \fi} 
\fi 

\usepackage{amscd} 
\usepackage{graphicx}  
\usepackage{amsmath,amssymb,array,calc,rotating} 
 
\ifnum\draftcontrol=1 
\fi 
 
\renewcommand\baselinestretch{1.25} 
\renewcommand\section{\@startsection {section}{1}{\z@}%
                                   {-3.5ex \@plus -1ex \@minus -.2ex}%
                                   {2.3ex \@plus.2ex}%
                                   {\normalfont\large\bfseries}} 
\renewcommand\subsection{\@startsection{subsection}{2}{\z@}%
                                   {-3.25ex\@plus -1ex \@minus -.2ex}%
                                   {1.5ex \@plus .2ex}%
                                   {\normalfont\normalsize\bfseries}} 
\renewcommand\subsubsection{\@startsection{subsubsection}{3}{\z@}%
                                   {-3.25ex\@plus -1ex \@minus -.2ex}%
                                   {1.5ex \@plus .2ex}%
                                   {\normalfont\normalsize\it}} 
\renewcommand\paragraph{\@startsection{paragraph}{4}{\z@}%
                                   {-3.25ex\@plus -1ex \@minus -.2ex}%
                                   {1.5ex \@plus .2ex}%
                                   {\normalfont\normalsize\bf}} 
 
\def\revise#1       {\raisebox{-0em}{\rule{3pt}{1em}}%
                     \marginpar{\raisebox{.5em}{\vrule width3pt\ 
                     \vrule width0pt height 0pt depth0.5em 
                     \hbox to 0cm{\hspace{0cm}{%
                     \parbox[t]{4em}{\raggedright\footnotesize{#1}}}\hss}}}}

\def\del          {\partial} 
 
\def\ee           {{\rm e}}

\def\tr           {\mathop{\rm Tr}}

\def\de#1#2{{\rm d}^{#1}\!#2\,}

\def\sqr#1#2{{\vcenter{\vbox{\hrule height.#2pt   
 \hbox{\vrule width.#2pt height#1pt \kern#1pt 
 \vrule width.#2pt}\hrule height.#2pt}}}}

\def\a{\alpha} 
\def\b{\beta} 
\def\r{\rho}

\def\la{\lambda} 
 
\def\be{\begin{equation}} 
\def\ee{\end{equation}} 
 
\def\m{\mu} 
\def\g{\gamma} 
\def\l{\lambda} 
\def\n{\nu} 

\catcode`\@=12 
 
\begin{document} 
 

\def\f{\frac} 
\def\nn{\nonumber} 
\def\w{\wedge}

\newcommand{\R}[1]{\mathbb{R}^{#1}} 
 
    
\def \el {{\ell}}    
\def \KK {{\cal  K}}    
\def \K {{\rm K}}    
\def \tz{\tilde{z}}    
    
\def \ci {\cite}    
\newcommand{\rf}[1]{(\ref{#1})}    
\def \la {\label}    
\def \const {{\rm const}}

\def \ov {\over}    
\def \ha {\textstyle { 1\ov 2}}    
\def \we { \wedge}    
\def \P { \Phi} \def\ep {\epsilon}    
\def \ab {{A^2 \ov B^2}}    
\def \ba {{B^2 \ov A^2}}    
\def \tv   {{1 \ov 12}}    
\def \go { g_1}\def \gd { g_2}\def \gt { g_3}    
\def \gc { g_4}\def \gp { g_5}\def \F {{\cal F}}    
\def \del { \partial}    
\def \t {\theta}    
\def \p {\phi}    
\def \ep {\epsilon}    
\def \te {\tilde \epsilon}    
\def \ps {\psi}    
\def \x {{x_{11}}}

\def\br{\bar{\rho}}    
\newcounter{subequation}[equation]

\def\pa{\partial}    
\def\e{\epsilon}    
\def\rt{\rightarrow}    
\def\tr{{\tilde\rho}}    
\newcommand{\eel}[1]{\label{#1}\end{equation}}    
\newcommand{\bea}{\begin{eqnarray}}    
\newcommand{\eea}{\end{eqnarray}}    
\newcommand{\eeal}[1]{\label{#1}\end{eqnarray}}    
\newcommand{\LL}{e^{2\lambda(r)}}    
\newcommand{\NN}{e^{2\nu(r)}}    
\newcommand{\PP}{e^{-2\phi(r)}}    
\newcommand{\non}{\nonumber \\}    
\newcommand{\CR}{\non\cr}

\makeatletter    
    
\def\thesubequation{\theequation\@alph\c@subequation}    
\def\@subeqnnum{{\rm (\thesubequation)}}    
\def\slabel#1{\@bsphack\if@filesw {\let\thepage\relax    
   \xdef\@gtempa{\write\@auxout{\string    
      \newlabel{#1}{{\thesubequation}{\thepage}}}}}\@gtempa    
   \if@nobreak \ifvmode\nobreak\fi\fi\fi\@esphack}    
\def\subeqnarray{\stepcounter{equation}    
\let\@currentlabel=\theequation\global\c@subequation\@ne    
\global\@eqnswtrue \global\@eqcnt\z@\tabskip\@centering\let\\=\@subeqncr    
    
$$\halign to \displaywidth\bgroup\@eqnsel\hskip\@centering    
  $\displaystyle\tabskip\z@{##}$&\global\@eqcnt\@ne    
  \hskip 2\arraycolsep \hfil${##}$\hfil    
  &\global\@eqcnt\tw@ \hskip 2\arraycolsep    
  $\displaystyle\tabskip\z@{##}$\hfil    
   \tabskip\@centering&\llap{##}\tabskip\z@\cr}    
\def\endsubeqnarray{\@@subeqncr\egroup    
                     $$\global\@ignoretrue}    
\def\@subeqncr{{\ifnum0=`}\fi\@ifstar{\global\@eqpen\@M    
    \@ysubeqncr}{\global\@eqpen\interdisplaylinepenalty \@ysubeqncr}}    
\def\@ysubeqncr{\@ifnextchar [{\@xsubeqncr}{\@xsubeqncr[\z@]}}    
\def\@xsubeqncr[#1]{\ifnum0=`{\fi}\@@subeqncr    
   \noalign{\penalty\@eqpen\vskip\jot\vskip #1\relax}}    
\def\@@subeqncr{\let\@tempa\relax    
    \ifcase\@eqcnt \def\@tempa{& & &}\or \def\@tempa{& &}    
      \else \def\@tempa{&}\fi    
     \@tempa \if@eqnsw\@subeqnnum\refstepcounter{subequation}\fi    
     \global\@eqnswtrue\global\@eqcnt\z@\cr}    
\let\@ssubeqncr=\@subeqncr    
\@namedef{subeqnarray*}{\def\@subeqncr{\nonumber\@ssubeqncr}\subeqnarray}    
    
\@namedef{endsubeqnarray*}{\global\advance\c@equation\m@ne    
                           \nonumber\endsubeqnarray}    
    
\makeatletter \@addtoreset{equation}{section} \makeatother    
\renewcommand{\theequation}{\thesection.\arabic{equation}}    
    
\def \ci {\cite}    
\def \la {\label}    
\def \const {{\rm const}}    
\catcode`\@=11    
    
\newcount\hour    
\newcount\minute    
\newtoks\amorpm \hour=\time\divide\hour by 60\minute    
=\time{\multiply\hour by 60 \global\advance\minute by-\hour}    
\edef\standardtime{{\ifnum\hour<12 \global\amorpm={am}    
        \else\global\amorpm={pm}\advance\hour by-12 \fi    
        \ifnum\hour=0 \hour=12 \fi    
        \number\hour:\ifnum\minute<10    
        0\fi\number\minute\the\amorpm}}    
\edef\militarytime{\number\hour:\ifnum\minute<10 0\fi\number\minute}    
    
\def\draftlabel#1{{\@bsphack\if@filesw {\let\thepage\relax    
   \xdef\@gtempa{\write\@auxout{\string    
      \newlabel{#1}{{\@currentlabel}{\thepage}}}}}\@gtempa    
   \if@nobreak \ifvmode\nobreak\fi\fi\fi\@esphack}    
        \gdef\@eqnlabel{#1}}    
\def\@eqnlabel{}    
\def\@vacuum{}    
\def\marginnote#1{}    
\def\draftmarginnote#1{\marginpar{\raggedright\scriptsize\tt#1}}    
\overfullrule=0pt    
    
 \def \lc {light-cone\ }    
    
\def\draft{    
        \pagestyle{plain}    
        \overfullrule=2pt    
        \oddsidemargin -.5truein    
        \def\@oddhead{\sl \phantom{\today\quad\militarytime} \hfil    
        \smash{\Large\sl DRAFT} \hfil \today\quad\militarytime}    
        \let\@evenhead\@oddhead    
        \let\label=\draftlabel    
        \let\marginnote=\draftmarginnote    
        \def\ps@empty{\let\@mkboth\@gobbletwo    
        \def\@oddfoot{\hfil \smash{\Large\sl DRAFT} \hfil}    
        \let\@evenfoot\@oddhead}    
    
\def\@eqnnum{(\theequation)\rlap{\kern\marginparsep\tt\@eqnlabel}    
        \global\let\@eqnlabel\@vacuum}  }    
    
\renewcommand{\rf}[1]{(\ref{#1})}    
\renewcommand{\theequation}{\thesection.\arabic{equation}}    
\renewcommand{\thefootnote}{\fnsymbol{footnote}}    
    
\newcommand{\newsection}{    
\setcounter{equation}{0}    
\section}    
    
\textheight = 22truecm     
\textwidth = 17truecm     
\hoffset = -1.3truecm     
\voffset =-1truecm    
    
\def \tx {\textstyle}    
\def \tix{\tilde{x}}    
\def \bi{\bibitem}    
    
\def \ov {\over}    
\def \ha {\textstyle { 1\ov 2}}    
\def \we { \wedge}    
\def \P { \Phi} \def\ep {\epsilon}    
\def \ab {{A^2 \ov B^2}}    
\def \ba {{B^2 \ov A^2}}    
\def \tv   {{1 \ov 12}}    
\def \go { g_1}\def \gd { g_2}\def \gt { g_3}    
\def \gc { g_4}\def \gp {    
g_5}    
\def \F {{\cal F}}    
\def \del { \partial}    
\def \t {\theta}    
\def \p {\phi}    
\def \ep {\epsilon}    
\def \ps {\psi}

\def \LL{{\cal L}}    
\def\o{\omega}    
\def\O{\Omega}    
\def\e{\epsilon}    
\def\pd{\partial}    
\def\pdz{\partial_{\bar{z}}}    
\def\bz{\bar{z}}    
\def\e{\epsilon}    
\def\m{\mu}    
\def\n{\nu}    
\def\a{\alpha}    
\def\b{\beta}    
\def\g{\gamma}    
\def\G{\Gamma}    
\def\d{\delta}    
\def\r{\rho}    
\def\bx{\bar{x}}    
\def\by{\bar{y}}    
\def\bm{\bar{m}}    
\def\bn{\bar{n}}    
\def\s{\sigma}    
\def\na{\nabla}    
\def\D{\Delta}    
\def\l{\lambda}    
\def\te{\theta} \def \t {\theta}    
\def\ta {\tau}    
\def\na{\bigtriangledown}    
\def\p{\phi}    
\def\L{\Lambda}    
\def\hR{\hat R}    
\def\ch{{\cal H}}    
\def\ep{\epsilon}    
\def\bj{\bar{J}}    
\def \foot{ \footnote}    
\def\be{\begin{equation}}    
\def\ee{\end{equation}}    
\def \P {\Phi}    
\def\un{\underline{n}}    
\def\ur{\underline{r}}    
\def\um{\underline{m}}    
\def \ci {\cite}    
\def \g {\gamma}    
\def \G {\Gamma}    
\def \k {\kappa}    
\def \l {\lambda}    
\def \L {{L}}    
\def \Tr {{\rm Tr}}    
\def\apr{{A'}}    
\def \m {\mu}    
\def \n {\nu}    
\def \W{{\cal W}}    
\def \eps {\epsilon}    
\def \ha{{    
 { 1 \ov 2}} }    
\def \de{{    
{ 1 \ov 9}} }    
\def \si{{    
 { 1 \ov 6}} }    
\def \fo{{    
{ 1 \ov 4}} }    
\def \ei{{    
{ 1 \ov 8}} }    
\def \rt {{\tx { \ta \ov 2}}}    
\def \rr {{\bar \rho}}    
    
\def\D{\Delta}    
\def\l{\lambda}    
\def\L{\Lambda}    
\def\te{\theta}    
\def\g{\gamma}    
\def\Te{\Theta}    
\def\tw{\tilde{w}}

\def\sn{\rm sn}  
\def\cn{\rm cn}   
\def\dn{\rm dn}

\def\hzero{\hat{0}} 
\def\ha{\hat{a}} 
\def\hb{\hat{b}} 
\def\hc{\hat{c}} 
\def\hd{\hat{d}} 
\def\he{\hat{e}}

\def\hone{\hat{1}} 
\def\htwo{\hat{2}} 
\def\hthree{\hat{3}} 
\def\hz{\hat{z}} 
\def\hteone{\hat{\theta}_1} 
\def\htetwo{\hat{\theta}_2} 
\def\hpone{\hat{\phi}_1} 
\def\hptwo{\hat{\phi}_2} 
\def\hpsi{\hat{\psi}}


\topmargin=0.50in   
    
\date{}    
    
\begin{titlepage}    
    
\version\versionno

\hfill MCTP-07-23\\    
    
\begin{center}    
    
{\Large\bf Black Holes in Cascading Theories:  }\\

\vskip .7 cm  
{\Large \bf  Confinement/Deconfinement Transition  } 
 
\vskip .7cm  
{\Large \bf and other Thermal Properties}

\vskip .7 cm    
{\large Manavendra Mahato${}^1$, Leopoldo A. Pando Zayas$^{1}$} 
 
\vskip .4cm  
 
{\large and C\'esar A. Terrero-Escalante$^{2}$} 
\vskip 1 cm

\end{center}

\vskip .4cm  
\centerline{\it ${}^1$ Michigan Center for Theoretical Physics}    
\centerline{ \it Randall Laboratory of Physics, The University of Michigan}    
\centerline{\it Ann Arbor, MI 48109-1120, USA}     
    
\vskip .4cm  
 
\centerline{\it ${}^2$  Departamento de F\'isica Te\'orica, Instituto de F\' isica}    
\centerline{ \it Universidade do Estado do Rio de Janeiro}    
\centerline{\it Maracan\~a, 20559-900 RJ, Brazil }


\vskip 1.5 cm    
    
\begin{abstract}    
We present numerical evidence for a transition between the 
Klebanov-Strassler background and a solution describing a black hole 
in the class of cascading solutions in the chirally restored phase. We 
also present a number of properties of this solution, including the 
running of the coupling constant, the viscosity to entropy ratio and the drag force on 
a quark moving in this background. 
\end{abstract}

\end{titlepage}    
\setcounter{page}{1} \renewcommand{\thefootnote}{\arabic{footnote}}    
\setcounter{footnote}{0}    
    
\def \N{{\cal N}}     
\def \ov {\over}

\section{Introduction}   
 
The AdS/CFT correspondence provides an alternative way to study the 
strongly coupled regime of gauge theories via  gravity duals. The 
original statement of the AdS/CFT correspondence identifies ${\cal 
N}=4$ supersymmetric  Yang-Mills with  IIB string theory on $AdS_5\times 
S^5$ \cite{malda,agmoo}. At finite temperature the gravity side  is 
described by nonextremal D3 branes and  the qualitative matching of 
the properties was one of the key observations in the understanding 
and eventual formulation  of the AdS/CFT correspondence 
\cite{igortemp}. Other interesting aspects of finite temperature 
theories, as seen  by the AdS/CFT correspondence  were  discussed by 
Witten \cite{wittenhp}. In particular, the Hawking-Page phase 
transition in the gravity side was related to the 
confinement/deconfinement transition on the field theory side. 
 
In a remarkable series of papers \cite{kw,gk,kn,kt,ks} Klebanov and 
collaborators carried out a program that concluded with a 
supergravity background that is dual to a 
confining field theory with chiral symmetry breaking. This background is known as the 
warped deformed conifold or the Klebanov-Strassler (KS) solution. In the context 
of the gauge/gravity correspondence it is natural to ask about the 
possible deconfinement transition for the dual field theory at finite 
temperature. The finite temperature phase of such theory is certainly 
very  interesting and has been tackled in various papers including 
\cite{alex,105,172}. In particular, reference \cite{172} constructed  
a perturbative solution whose regime of validity is restricted to 
high temperature and small value of the $F_3$ flux: $\int_{\Sigma_3} 
F_3 =P$. Further improvements to this solution were presented in 
\cite{aby}.  Knowing the solution only asymptotically in the radial  
coordinate and for a specific regime of parameters prevents us from 
extracting the full thermodynamics and from being able to understand 
possible phase transitions.  
In particular, in order to answer questions that require knowing the solution for all values of the radial coordinate 
(for instance, those questions involving the action),  
it is important to have the full 
solution which was found numerically in \cite{PT}. 

The thermodynamic aspects of strongly coupled field theories have gained a lot of attention recently, 
largely stimulated by RHIC. Some of the thermodynamic aspects discussed in this paper 
have been considered in 
simpler supergravity backgrounds in 
\cite{cobitemp,temp1,temp2,temp3,temp4}. 
The literature is by now very extensive, a representative sample of the 
different questions that are being pursued can be found 
in the pages of the Perimeter Institute, where part of our results were preliminarily presented 
\cite{PI}.
 
The paper is organized as follows. 
We review the construction of the cascading black hole in section \ref{sec:review}. In section 
\ref{sec:hptransition} 
we present our main result which is the transition between the 
KS solution and the cascading black hole. 
Section \ref{sec:properties} discusses various properties of the 
cascading black hole, including the running of the coupling constant at finite temperature, comments on the viscosity bound 
and the drag force. 
We conclude in section \ref{sec:conclusions} with a brief discussion of our main results and some open questions.

\section{Review of the cascading black hole}\label{sec:review} 
To understand the Ansatz for the finite temperature solution we 
start from the conformal case where the supergravity 
background is $AdS_5\times T^{1,1}$ and the field theory was discussed 
in \cite{kw}. The  five dimensional  manifold denoted by  $T^{1,1}$ 
is parametrized by coordinates $(\psi, \theta_1, 
\phi_1,\theta_2,\phi_2)$.  
The construction of the Ansatz follows directly the one presented 
originally in 
\cite{105}.  
For the metric we consider a generalization consistent with the $U(1)$ 
symmetry generated by $\psi$-rotations\footnote{In the gauge theory 
\cite{kw} this symmetry is identified with  the $U(1)_R$. Restoring 
this symmetry at high temperature is understood as chiral  symmetry 
restoration \cite{172}. Therefore, the solution we constructed in 
\cite{PT} corresponds to the  deconfined phase due to the 
presence of a horizon but also to the  phase with the chiral symmetry 
restored due to the fact that $U(1)_\psi$ is  a symmetry of the 
background.}.  The Ansatz in question depends on four  functions 
$(x,y,z,w)$   of the radial coordinate denoted by $u$: 
 
\be 
\eqlabel{metricPT} 
ds^2 =  e^{2z} ( -e^{-6x} dX_0^2 + e^{2x} dX_i dX^i) + e^{-2z}  ds^2_6 
\ , 
\ee 
where 
\bea 
ds^2_6 &= & e^{10y} du^2 + e^{2y} (dM_5)^2  \ ,  \nonumber \\ (dM_5)^2 
& = & e^{ -8w}  e_{\psi}^2 +  e^{ 2w} 
\big(e_{\theta_1}^2+e_{\phi_1}^2 + 
e_{\theta_2}^2+e_{\phi_2}^2\big) \equiv e^{ 2w} ds_5^2. 
\eea 
The Funfbein is: 
\be 
 e_{\psi} =  {\frac13} (d\psi +  \cos \theta_1 d\phi_1  +  \cos 
 \theta_2 d\phi_2)  \  ,  \quad  e_{\theta_i}=\frac1{\sqrt 6} 
 d\theta_i\ ,  \quad  e_{\phi_i}= 
\frac1{\sqrt 6} \sin\theta_id\phi_i \ . 
\ee 
The qualitative meaning of the metric 
functions $x,y,w$ and $z$ was explained in \cite{105,PT}.  
 
The Ansatz for the $p$-form  fields is  as in the original Klebanov-Tseytlin (KT)  solution \cite{kt}: 
 
\bea 
F_3 &=& \   P e_\psi \wedge ( e_{\theta_1} \wedge e_{\phi_1} - 
e_{\theta_2} \wedge e_{\phi_2})\ , \nonumber \\  B_2  &=& \    f(u)  ( 
e_{\theta_1} \wedge e_{\phi_1} -  e_{\theta_2} \wedge e_{\phi_2}) \ , 
\nonumber \label{eq:forms} \\ F_5&=& {\cal F}+* {\cal F}\ , \quad  \ \ \ \ {\cal F} = 
K(u)  e_{\psi}\wedge e_{\te_1} \wedge e_{\p_1} \wedge e_{\te_2}\wedge 
e_{\p_2}\ . 
\eea 
Note that the form of $F_3$ is such that it describes a  constant flux 
along a 3-cycle, that is, $\int_{\Sigma_3}F_3=P$.  In some other 
notation this is called the number of  fractional D3 branes. The 
Bianchi identity for the 5-form, \ $ d*F_5=dF_5=H_{3}\wedge F_3$, 
implies 
\be 
K (u)  = Q + 2 P f (u) \, , 
\ee 
where $Q$ is a constant.
$P$ and $Q$ are known as $M$ and $N$ in the standard literature 
(see review \cite{hko}).

Thus, in the presence of 3-form flux $(P\ne 0)$, the flux of $F_5$ 
varies with the radius.  The fact that $K(u)$ depends on the 
coordinate $u$ is very novel and has  interesting physical 
implications. Phenomenologically, a very attractive property of  
this class of supergravity solutions is that it encodes the logarithmic running  
of a combination of gauge couplings in field theory. It does so via a varying $B_2$  
field which is compensated by a constant $F_3$ flux through a  
3-cycle.  The  
five-form, which is constant in most solutions, varies according to the Bianchi identity  
$dF_5=H_3\wedge F_3$ and generates a varying flux as  
$\int_{\Sigma_5} F_5$ depends on the radial coordinate. The supergravity solution therefore  
has varying flux. This varying flux was interpreted in \cite{ks} as 
the dual of a Seiberg duality cascade, coining therefore the term of cascading 
solutions.

\subsection{The system of equations from reduction to 1-d} 
An efficient way to derive the system of type IIB supergravity 
  equations of motion  is to  start with the 1-d effective action  for 
  the radial evolution which follows from the 10-d action. This approach is 
  convenient because to establish the transition we compare the 
  actions. So, a simple action is an advantage.

The type IIB supergravity equations of motion follow from the action 
$$ 
 S_{ 10} =-\frac 1{2\kappa_{10}^2}  \int d^{10} x \bigg( \sqrt{-g_{10}} 
\bigg[ \ R_{10} 
 - { \textstyle \frac 12} (\partial \Phi)^2 - { \textstyle \frac 1{12}} e^{-\Phi}   (\partial B_2)^2  $$ \be - \ { \textstyle \frac 12} 
e^{2 \Phi} (\partial {\cal C})^2 - { \textstyle \frac 1{12}} e^{ \Phi} 
(\partial C_2  - {\cal C} 
\partial B_2) ^2 
- { \textstyle \frac 1 {4\cdot 5!}}  F^2_5\ \bigg] 
{ \textstyle \frac 1{ - 2\cdot 4! \cdot (3!)^2    }} {\ep_{10}} C_4 \partial C_2 \partial 
- B_2 + ... \bigg) \ , \la{acto} \ee 
 $$(\partial B_2)_{...} =3 \partial_{[.} B_{..]} ,\ \  \ 
\   (\partial C_4)_{....} \equiv 5  
\partial_{[.} C_{....]}  , \ \ 
F_5= \partial C_4 + {5} (B_2 \partial C_2 - C_2 \partial B_2) , $$ 
 supplemented with the on-shell constraint $F_5=*F_5$.

For the metric (\ref{metricPT}) 
$$ \sqrt G= \frac 1{108}  e^{10y -2 z}\sin\theta_1\sin\theta_2, $$ 
 
\be 
\int d^{10} x \sqrt G  R\ \  \to \ \  \ \   
 C_{b.h.}\frac 1{27}\int du \big[ 5 y'^2 - 3 x'^2 - 2 z'^2  - 5 w'^2 + 
e^{8y} ( 6 e^{-2w} - e^{-12 w} )  \big] \ , 
\label{eq:PTaction1}
\ee 
where $C_{b.h.}=64\, \pi^3\, \beta_{b.h.} \,V_3$. The inverse 
temperature of the black hole is represented by $\beta_{b.h.}$ and 
$V_3$ is the spatial volume in the $x^1,x^2$ and $x^3$ directions.

The matter part is  
$$\int d^{10} x \sqrt G \ [ - \frac 12 ( \del\P)^2 +...] \ \  \to \ 
\  $$ 
 
\be 
  - C_{b.h.}\frac 1{27}\int du \    \frac 18 \bigg[  \Phi'^2 + 2 
 e^{-\Phi + 4z -4y - 4 w} f'^2  + 2  e^{\Phi + 4z + 4y + 4w} P^2 + 
 e^{8z} (Q + 2 P f)^2   \bigg]\ . 
\label{eq:PTaction2}
\ee 
The 1-d 
effective lagrangian is: 
$$ 
L=   T  - V \  , 
$$  
 
\be 
T =  5 y'^2  -  3 x'^2 - 2 z'^2  - 5 w'^2  - \frac 18 \P'^2 
- \frac 14  e^{-\P +  4z -4y - 4 w } f'^2 
\ ,  
\ee 
\be 
\label{lagg}   
 V = -    e^{8y} ( 6 e^{-2w} - e^{-12 w} ) 
+  \frac 14 e^{\P+  4z + 
 4y + 4 w } P^2 +  \frac 18  e^{8z} (Q + 2 P f)^2 \ , 
\ee 
supplemented with  the ``zero-energy"  constraint $T+V=0$.

This is effectively a classical mechanical system.  The simplest equation is 
 for the nonextremality function $x$ (note that it breaks Loretz 
 invariance in the $(X^0,X^1,X^2,X^3)$ plane): 
\be 
\eqlabel{eq:x} 
x''=0 \ , \ \ \ \  x= a u \ , \ \ \ \     a=\const . 
\ee 
The reason for such a simple equation is that, as explained in 
\cite{105},  it does not appear in the effective one dimensional 
Lagrangian, except for its kinetic term. 
 
The other functions $y,w,z,f$ and $\P$ should satisfy a 
coupled system of equations: 
\bea 
\label{eq:sys} 
10y'' - 8 e^{8y} (6 e^{-2w} - e^{-12 w})   + \P'' &=&0, \nonumber \\ 
10w'' - 12 e^{8y} ( e^{-2w} - e^{-12 w})   - \P'' &=&0, \nonumber \\ 
\P''    + e^{-\P + 4z - 4y-4w} (f'^2 -  e^{2 \P + 8 y+8w} P^2)&=&0, \nonumber\\ 
4z'' -  (Q+ 2 P f)^2  e^{8z} - e^{-\P + 4z - 4y-4w} ( f'^2 +  e^{2 \P 
+ 8 y+8w} P^2) &=&0, \nonumber \\   (e^{-\P + 4z - 4y-4w} f')' - P (Q+ 
2 P f) e^{8z} &=&0. 
\eea 
The solutions to this system are  subject  to the  zero-energy constraint 
$T+V=0$, i.e. 
\bea 
\label{eq:constraint} 
&&5  y'^2   - 2 z'^2  - 5 w'^2 - {\frac18} \P'^2  - {\frac14}  e^{-\P 
+  4z -4y - 4 w } f'^2  -   \  e^{8y} ( 6 e^{-2w} - e^{-12 w} ) 
\nonumber\\ && +  {\frac14} e^{\P+  4z + 4y + 4 w } P^2 +  {\frac18} 
e^{8z} (Q + 2 P f)^2   - 3 a^2 = 0 \ . 
\eea 
This constraint was used in \cite{PT} as a criterion for quality of 
the numerical solution and it was established that for various known 
solutions it was of the order $10^{-16}$ where the reliable accuracy was shown to 
be $10^{-10}$ 
(see fig.\ref{fig:constraint} and 
section \ref{sec:numerics} for a detailed description).  
 
The function $y$ amounts to a choice of the radial coordinate. This is 
relevant for understanding the dimensions of all quantities. Note that 
in particular, the system (\ref{eq:sys}) and the  constraint 
(\ref{eq:constraint}) are invariant under  $e^y\to L_0 e^y$ and $u\to 
L_0^{-4}u$ if we assume that $Q\to L_0^4Q, P\to L_0^2P, a\to L_0^4a$ 
and $f\to L_0^2f$.  We therefore express all dimensionfull quantities 
in units where $L_0=1$.

{\it Standard regular non-extremal D3-brane solution:} 
To develop our intuition, we present the nonextremal D3 brane of the system (\ref{eq:sys})  
in the radial $u$-coordinate. It takes the following form: 
\be 
\eqlabel{eq:regd3} 
e^{4y} =  \frac{ a}{ \sinh 4a u  } \ , \ \ \ \ \ \ 
 e^{4z} =  \frac{ a}{  q \sinh 4 a u  } \ ,\ \ \ \ \ \ \ 
e^{4x} = e^{4au} \,   .
\ee 
Note that  near the horizon ($u\to \infty$)  
\be 
y= y_* - a u +  \frac 14  e^{-8au} + O( e^{-16au}) \ , \ \ \ \ \ 
z= z_*  - a u + \frac 14  e^{-8au} + O( e^{-16au}) \ , 
\ee 
\be 
 y_* = \frac 14 \ln 2 a \ , \ \ \ \  
z_* = \frac 14 \ln \frac{2 a}{ q} \ , \ \ \ 
\ee 
A more recognizable form of this solution is given by: 
\be 
ds^2 =  h^{-1/2} ( g dX_0^2 + dX_i dX_i) 
+  h^{1/2}  [ g^{-1}  d\r^2  
+ \r^2 (d M_5)^2] \ , \ \ \ \  
\ee 
\be 
g= e^{-8x} = 1 - \frac{2 a}{  \r^{4}}  \  ,  
 \ \ \ \ \ \ \  
  \r^4= \frac{ 2 a }{1- e^{-8a u} } \ , 
\ \ \ \ \  h=  e^{-4z- 4x}= \frac{ q }{  \r^{4} }  \ .  
\ee

Let us clarify the relationship between the radial coordinate $u$ and the more  
standard coordinate $\rho$. Using $2a=\rho_0^4$,  we have 
\be 
\eqlabel{eq:urho} 
du =\frac{d\rho}{\rho^5}\left(1-\frac{\rho_0^4}{\rho^4}\right)^{-1}. 
\ee 
This can be integrated to  
\be 
\eqlabel{eq:urhohor} 
u=-\frac{1}{4\rho_0^4}\ln \left(1-\frac{\rho_0^4}{\rho^4}\right). 
\ee 
Note that in the domain of $\rho_0 \le \rho < \infty$ we have that $u$ 
ranges in $0 <u<\infty$. For large values of $\rho$ we have that 
$u\approx 1/4\rho^4$.  
The position of the horizon which is finite in the $\rho$ coordinate becomes infinite in  
the $u$ coordinates.

\subsection{Universality of regular nonextremal D3 brane horizons} 
 
For the general solution of the system (\ref{eq:sys}), in the 
$u$-coordinates the area of a surface defined by a horizon at $u={\rm constant}$ is 
\be 
A=V\o_5 \exp\left(-2z+3x+5y\right). 
\label{eq:horA} 
\ee 
Given that the equation of motion for $x$ has the general solution 
$x=a u$ we are  
forced into the following situation.  
If the horizon is at $u\to \infty$, as is the case for the non-extremal D3 brane solution,  
 then  for the area $A$ to 
be finite we  
need the following asymptotics for  
$z$ and $y$: 
\be 
z\to \alpha\, a u + z_*, \qquad y\to \beta\, a u + y_*,  
\eqlabel{eq:yzasym} 
\ee 
with the condition that  
\begin{equation} 
3-2\alpha+5\beta=0 \, .
\eqlabel{eq:yzcond} 
\end{equation} 
Note that the regular nonextremal D3 brane corresponds to $\a=\b=-1$.  
The main claim is that: {\it The existence of a regular horizon fixes 
the asymptotic  
behavior of the metric coordinates 
$x,y$ and $z$ near the horizon.} 
Eq.(\ref{eq:yzcond}) was used in \cite{PT} as a criterion of quality
of the numerical solution near the horizon (see table \ref{tab:figpars} for the precision of this 
criterion for the solutions we discuss here).
 
Similarly, one can obtain an expression for the temperature. Namely,  the relevant part of the metric is  
\be 
ds^2=e^{2z-6x}d\tau^2+e^{-2z+10y}du^2. 
\ee 
We introduce a new radial coordinate as: 
\be 
\r=e^{z-3x}. 
\ee 
We can now rewrite the metric as: 
\be 
ds^2=\frac{e^{-4z+10y+6x}}{(z'-3x')^2}\bigg[d\r^2+\r^2\left(e^{2z-5y-3x}(z'-3a)\,\, d\tau\right)^2\bigg]. 
\ee 
Note that, again, {\it the requirement of finite temperature fixes the large-$u$ asymptotic of various metric functions to  
be $2z-5y-3x \to {\rm constant.}$} Requiring the absence of conical singularity 
 we find that the temperature defined as the inverse of the period is 
\be 
T=\frac{|\a-3|}{2\pi}\,a\,e^{2z_*-5y_*}, 
\label{eq:BHT} 
\ee 
where we used that near the horizon the asymptotic form of $z$ and 
$y$ is given by (\ref{eq:yzasym}).

 
\subsection{Numerics} 
\label{sec:numerics} 
In  Ref.\cite{PT}  
we gave a detailed description of the construction of solutions to the system (\ref{eq:sys}) which 
contain regular black holes in the infrared (large values of $u$) and a cascading behavior in the 
ultraviolet (small values of $u$).  
We will summarize here our approach.
For more details on the numerical method  
and the understanding of the numerical output,  
we refer the interested reader to Ref.\cite{PT}. 

To solve the equations for the metric functions we use 
the seventh-eight order continuous Runge-Kutta method. 
Thanks to 
its adaptive scheme,  
this method provides a great control upon the output accuracy. 
We tested and fitted the numerical procedure  
using a number of known analytical solutions, 
some of them with fixed values of the parameters 
(for instance, the non-extremal D3 black hole, with $P=0$,  
and the KT background, with $a=0$). 
Analyzing the sensitivity of the solutions 
to the variation of these parameters, 
we found that for a method tolerance of $10^{-14}$,  
an error of $10^{-10}$ can be safely regarded to be negligible, 
what allowed us to set our numerical `zero' to this last value.  

Using the non-extremal D3 solution ($P=0$),  
we set the boundary conditions for our ten variables at  
the value of u corresponding to the $90\%$ of the distance to 
the horizon. 
A correction is introduced to enforce the solution with $P\neq 0$ 
to satisfy constraint (\ref{eq:constraint}) in this boundary.  

Next we integrate backward and, for a given value of $a$ and $Q$, 
we find the value of $P$ such that the space is  complete,  
i.e., $u_{sing}\approx 0$ (see section \ref{sec:completeness} and table \ref{tab:figpars}). 
We use 
\be
f_1+P{\rm e}^{\Phi_0+4y_0+4w_0}=0\, ,
\label{eq:UVcriterion} 
\ee
to check that this singularity is of the type  
given by 
\be 
z=-\frac14\ln \big[4(u-u_{sing})\big]\, , 
\label{eq:zUV}
\ee
and that the remaining fields are analytical at $u_{sing}$.
In this criterion, the subindices indicates that these are
coefficients in Taylor series for the
corresponding field at $u_{sing}$.

We then integrate the system forward and identify numerically the presence of a 
horizon.  
One of the criteria used is that given by Eq.(\ref{eq:yzcond}). 

In table \ref{tab:figpars} we present the values of the parameters  
and the results for the infrared and ultraviolet criteria of quality 
for four cases that are going to be used as examples throughout this 
manuscript. 
\begin{table}[t] 
\centering 
\begin{tabular}{|c|c|c|c|c|c|} 
\hline 
$a$&$P$&$u_{90\%}$&$3-2\alpha+5\beta$ 
&$u_{sing}$&$f_1+P{\rm e}^{\phi_0+4y_0+4w_0}$\\ 
\hline 
$1$&$4.26906025$&$2.19361453$&$-9\times 10^{-15}$ 
&$3\times 10^{-10}$&$0.00001257$\\ 
\hline 
$0.5$&$2.87343889$&$4.387229$&$-7\times 10^{-15}$ 
&$3\times 10^{-11}$&$0.0000065$\\ 
\hline 
$0.1$&$1.094389139$&$21.9361453$&$-2\times 10^{-15}$
&$4\times 10^{-10}$&$0.00000002$\\ 
\hline 
$0.01$&$0.316430028$&$219.361453$&$10^{-15}$
&$2\times 10^{-9}$&$0.0000000002$\\ 
\hline 
\end{tabular} 
\caption{Values of the parameters used for the four cases presented in this manuscript.  
$Q=L_p=1$ was used everywhere.} 
\label{tab:figpars} 
\end{table} 

\begin{figure}  
\centerline{\includegraphics[width=0.55\linewidth]{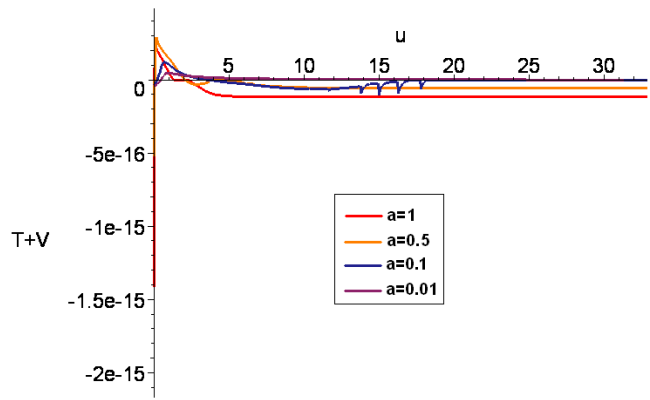}}  
\caption{The Hamiltonian constraint for the cases in table \ref{tab:figpars}.}   
\label{fig:constraint}  
\end{figure}

We also recall that 
system (\ref{eq:sys}) was modified in such a way that the solutions  
must automatically satisfy constraint (\ref{eq:constraint}). 
This allows to use the constraint as a dynamical criterion of the 
quality of the numerical output for any value of the parameters. 
Indeed, all our solutions are warranted to satisfy 
constraint (\ref{eq:constraint}) 
within the accuracy given by our numerical `zero'.
This can be verified in fig.\ref{fig:constraint}. 
For all the four cases in the table,
the $T+V=0$ constraint 
is satisfied in the whole range of the radial coordinate. 


\subsection{UV asymptotics and completeness of the solution}\label{sec:completeness} 
Let us first discuss the completeness of our solution.  
The statement that completeness of the space requires $u_{sing}=0$  
was made explicitly in \cite{PT} (see section 4.2 there), however, the details were not shown.  
Here we provide those details since they are crucial  
in clarifying the number of parameters of the solution\footnote{This point has been an important  
source of discussion with A. Buchel and we hope that this section clarifies that $u_{sing}=0$ fixes one of the  
three seemingly free parameters $(Q,P,a)$.}.    

The asymptotic form of the relevant part of the metric, 
as obtained in \cite{PT}, is given for $z(u)$ by Eq.(\ref{eq:zUV}) 
and for $y(u)$ by $\exp(4y)=1/4u$. 
The first expression was widely discussed in  
\cite{PT} and we refer the reader to that paper for details, 
including the fittings. The expression  
of $y$ is ubiquitous in the  
region of small values of $u$. 
It is valid in the asymptotic region for all the solutions we know: 
KW \cite{kw}, nonextremal D3 branes \cite{105}, KT \cite{kt} and KS \cite{ks}. 
The best fit analysis presented in \cite{PT} indicates 
that this seems to be also
the case for our numerical solution.
 
To discuss completeness we consider an outgoing null geodesic  
toward the asymptotic regions of small $u$. The corresponding effective Lagrangian is: 
\be 
{\cal L}=-e^{2z-6x}\dot{t}^2 + e^{-2z+10y}\dot{u}^2\, . 
\ee 
With this information the affine time for a geodesic to reach $u_{sing}$ satisfies: 
\be 
d\lambda =\frac{1}{E}u^{-5/4}du.
\ee 
Note that this is independent of the precise form of the $z$ function in the metric. The structure of the function  
$z$ enters only through the integration limits. Namely, the above equation can be integrated to yield 
\be 
\eqlabel{eq:completeness} 
\lambda= \frac{4}{E}\left(\frac{1}{u_{sing}}-\frac{1}{u_{in}}\right). 
\ee 
Here $u_{in}$ should be viewed as a cutoff, beyond which we do not know the precise formula for $y$. 
However,  
we assume that this geodesic originates somewhere near the black hole horizon. Note that the value $u_{sing}$ enters just through  
the integration limits, since the full solution does not exist for $u<u_{sing}$. As can be seen  
from (\ref{eq:completeness}) the  
affine parameter is infinite only for $u_{sing}=0$. We need an infinite affine parameter to guarantee completeness of the  
solution. Thus, {\it the solution is complete only for $u_{sing}=0$.}

\subsection*{The number of relevant parameters} 
 
Thus, superficially, our solutions depend on three parameters, 
$a$, $P$ and $Q$. 
However, the condition  
$u_{sing}(\cdots;a,P,Q)=0$ imposes a new constraint 
on the solutions that effectively reduces the number of parameters to two. 
 
\begin{figure}  
\centerline{\includegraphics[width=0.55\linewidth]{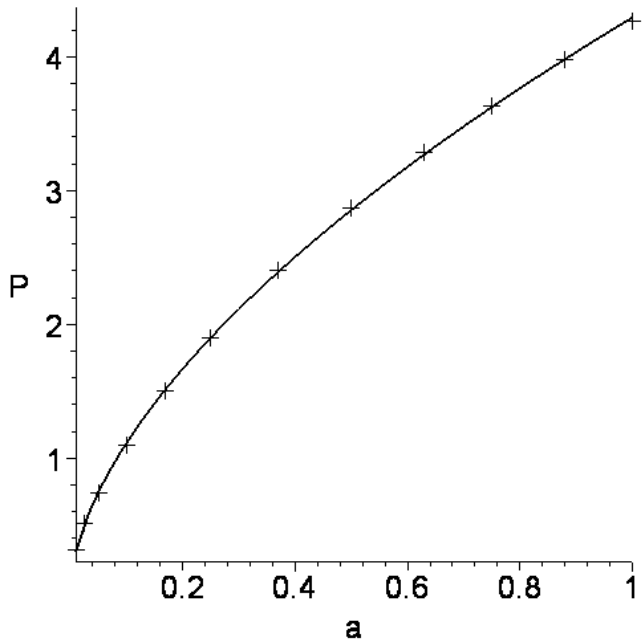}}  
\caption{Values of $P$ and $a$ that guarantee $u_{sing}$ to vanish for $Q=1$.} 
\label{fig:P_a} 
\end{figure}

For instance,  
fixing $Q=1$, for the values of $a$ given in table \ref{tab:figpars}, 
it is found that the expression, 
\be
\eqlabel{eq:using}
u_{sing}=b{\rm e}^{-P^2}+b_0 + b_1 P + b_2 P^2 + \cdots \, , 
\ee
fits very well the numerical output. 
Note that the first term is  
clearly related to the strong energy scale as indicated by the KT formula, 
that is, it is related to the  
radial position where the KT warp factor vanishes; of course, 
there are many corrections as the KT warp factor  
is not the full answer. To illustrate the origins and robustness of the expression (\ref{eq:using}), 
let us first discuss a specific point in the graph \ref{fig:P_a} 
to highlight  the importance of the  
first term in the above expansion. 
Namely,  for $a=0.5$, a best fit to the value of $u_{sing} $ without the exponential term (i.e., setting b=0) 
yields the following result for the coefficients $b_i$:
\begin{eqnarray} 
\{0.05854539, -0.25108172, 0.14454983, -0.03078276, \nonumber \\
0.00347435, -0.00020362,  0.488\times 10^{-5}\}     \nonumber \, ,
\end{eqnarray}
with a goodness of fit on the order of $10^{-7}$. It is important to stress that for each term we 
verify that $b_i \le b_{i+1}$, which indicates convergence.  
Now, introducing the exponential term, 
the result is $b=5.49169633$, 
\begin{eqnarray} 
\{-0.09506874,-0.09298412, 0.07882655, -0.01663906, \nonumber \\
 0.00180981, -0.00010186, 0.234\times 10^{-5}\}     \nonumber \, ,
\end{eqnarray} 
with a goodness of fit on the order of $10^{-9}$. 
The ratio $b_i/b_{i+1}$ is also much better behaved. 
In general the exponential term seems to be crucial for the form of $u_{sing}$. 
Thus, 
for $(a,Q)$ fixed,
by setting $u_{sing}=0$  
in expression (\ref{eq:using}), 
we are this way selecting a given value of $P$. 
A similar analysis was carried out for the other values of 
$a$ in table \ref{tab:figpars} 
and another ten cases not included in the table  
but shown as crosses in fig.\ref{fig:P_a}. 
The curve in this figure represents the pairs $(a,P)$ that guarantee $u_{sing}$ to vanish for $Q=1$.
It was obtained as the best fit solution,
\[
P= 4.26846923\,\, a^{0.59296839} + 0.02429978 
\]
with a goodness of fit of $0.00124$, 
which is satisfactory if the small amount of data used is taken into account.
 
Summarizing,  
for any values of two of the parameters determining our solution,  
 the value of the third one is fixed  
by imposing completeness of the solution in the asymptotically UV region.
 
\section{The Hawking-Page transition}\label{sec:hptransition} 

There is mounting evidence for the transition we are going to 
describe here. In particular, similar transitions have been established in some simpler 
holographic models, including 
\cite{herzogtransition,braziltransition,Cai:2007zw,Cai:2007bq}. 

In our case, 
we assume that the deconfined phase is dual to the cascading black hole reviewed in the previous section,
while the confined phase is dual to the warped deformed conifold, i.e., the Klebanov-Strassler background \cite{ks}.
The relevant expressions of this well known solution 
were compiled in the appendix
in terms of the coordinates and notation we use in this paper.

To accurately compute the actions we used an adaptive 3-5 Simpson's quadrature. 
 This implies $2^{n-1} 8^n$ evaluations of the Lagrangian, 
where $n$ is the number of iterations needed to  
reach the desired precision (set here to $10^{-4}$) for the integral.  
For the cascading black hole the Lagrangian 
is given in terms of the numerical solutions of system (\ref{eq:sys}). 
We gave priority to the output accuracy over the computational effort.  
This means being able to obtain the actual numerical solution  
of (\ref{eq:sys}) at every point required by the quadrature, 
instead of finding it by interpolating between  
previously calculated solutions at a given set of points. 
The whole computation of the cascading black hole  
action usually takes more than 12 hours in an ordinary desktop computer. 
Compared with that, estimating the  
action of the KS background is a quick task, 
since the evaluation of the corresponding Lagrangian only  
involves the numerical integration of (\ref{eq:KSI})
which for large $\tau$ can be accurately estimated using expression 
(\ref{eq:KSIuv}). 
Taking the above into account, 
we decided to compute the cascading  
black hole action 
for the four sets of parameters given in table \ref{tab:figpars} 
and compare the results  
with the output of the KS action as we vary $\epsilon$. 
 
\begin{figure}  
\centerline{\includegraphics[width=0.55\linewidth]{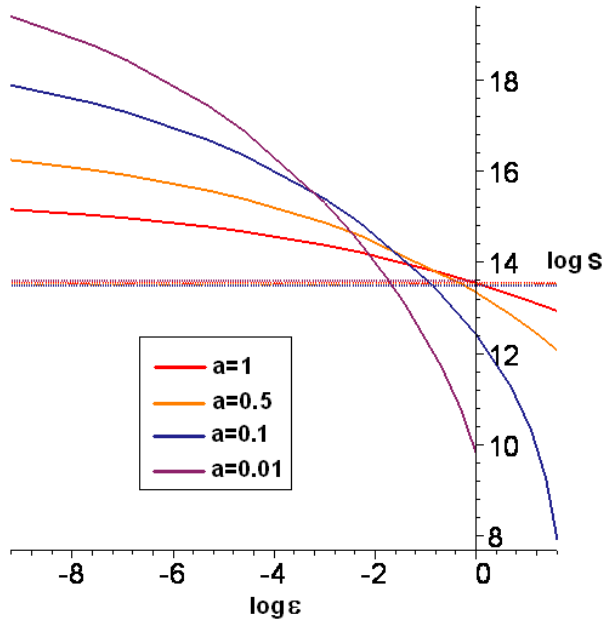}}  
\caption{The Hawking-Page transition. 
The horizontal lines represent the values of the cascading black hole action for the four cases analyzed in this paper. 
The curves show the $\epsilon$-dependence of the corresponding KS actions.}   
\label{fig:hp}  
\end{figure}

So, we calculated the cascading black hole action 
as given by Eqs.(\ref{eq:PTaction1}) and (\ref{eq:PTaction2}), 
without considering the numerical factors common to both actions.  
The integration was done between $u_R=10^{-8}$ and $u_{90\%}$. 
Following the order in table \ref{tab:figpars},  
the four results are $\{738451, 762361, 733711, 804269 \}$, 
and their logarithms are shown 
in fig.\ref{fig:hp} as horizontal lines. 
To leading order the asymptotic values of the action are independent of  $a$. This can be seen by directly 
evaluating the pieces of the action as given in (\ref{eq:PTaction1}) and (\ref{eq:PTaction2}). 
For example, the leading contribution from (\ref{eq:PTaction1}) comes from $z'{}^2$ and after the 
integration is proportional to $1/u_R$. Similarly, from the matter part (\ref{eq:PTaction2}), the leading 
contribution comes from the last term which is independent of $a$; this term also contributes a term 
proportional to $1/u_R$. Note that in terms of the standard radial coordinates, this corresponds to $R^4$ where 
$R$ is a UV cutoff radius. This is the typical volume divergent behavior of  $AdS_5$  as shown in 
\cite{wittenhp,herzogtransition,braziltransition}.

To compare these results with those for the KS action   
is necessary for both backgrounds to have the same  
physical temperature. This is implemented by requiring that the 
physical perimeter of the temporal directions match at the point of comparison: 
\begin{equation} 
\label{eq:matchtemp} 
\beta_{b.h.} e^{z(u_R)-3x(u_R)}=\beta_{KS}h^{-1/4}(\tau_R). 
\end{equation} 
 
This equation allows to determine $\beta_{KS}$, 
since for the black hole, 
$\beta_{b.h.}$ is determined by the absence of conical 
singularities (see equation (\ref{eq:BHT})). 
However, we still need a 
way to relate $\tau_R$ and $u_R$. 
As can be seen from various examples 
discussed in \cite{105,PT}, in the UV the coordinate $u$ is related to 
the standard radial coordinate as $u\approx 1/4r^4$; this is also true 
for the nonextremal D3 brane 
(see equation (\ref{eq:urhohor})) 
as well 
as for the KT solution, the b-deformed conifold  \cite{105}, 
KS and the resolved conifold \cite{resolved}. The relationship 
between the radial coordinate $\tau$ and the standard conifold radius 
is  
\begin{equation} 
dr=\frac{\epsilon^{2/3}}{2^{1/6}3^{1/2}}\frac{\sinh\tau}{(\sinh 2\tau - 
2\tau)^{1/3}}d\tau. 
\end{equation} 
In the limit of large values of $r$ we obtain: 
\begin{equation} 
r \approx \frac{3^{1/2}}{2^{5/6}}\epsilon^{2/3} e^{\tau/3}. 
\end{equation} 
 
Thus, in equation (\ref{eq:matchtemp}) we use that  
\begin{equation} 
\tau_R\approx  \frac14\ln(\epsilon^{-8}u_R^{-3})-0.954771252 + \delta\tau_R. 
\end{equation} 
The correction $\delta\tau_R$ is required to account  
for the limitations of the numerical analysis to deal with 
actual asymptotic values. 
It is a function of $a$ and $\epsilon$  
and we have found it to have the approximate form, 
\begin{equation} 
\label{eq:tau_corr} 
\delta\tau_R = C(a)\epsilon^{1/5}\, . 
\end{equation} 

Our strategy to establishing the existence of the transition is as follows.
We assume that the transition takes place at a value of  
$\epsilon$ such that $a^{-8/3}\epsilon=1$, then we find the $C(a)$ that makes both actions equal. 
Following the order in table \ref{tab:figpars}, we obtained that  
$C(a)=\{-2.27, -3.264, -5.85, -9.8095 \}$. 
We note that neither the factor $C(a)$ nor the power $1/5$ are solutions of a best fit problem 
but just values that give satisfactory results for all cases analyzed here. Our key observation then reduces to establish 
that the difference of actions changes sign appropriately around this point. 
 
\begin{figure}[htp]
\centering
\includegraphics[width=0.45\linewidth]{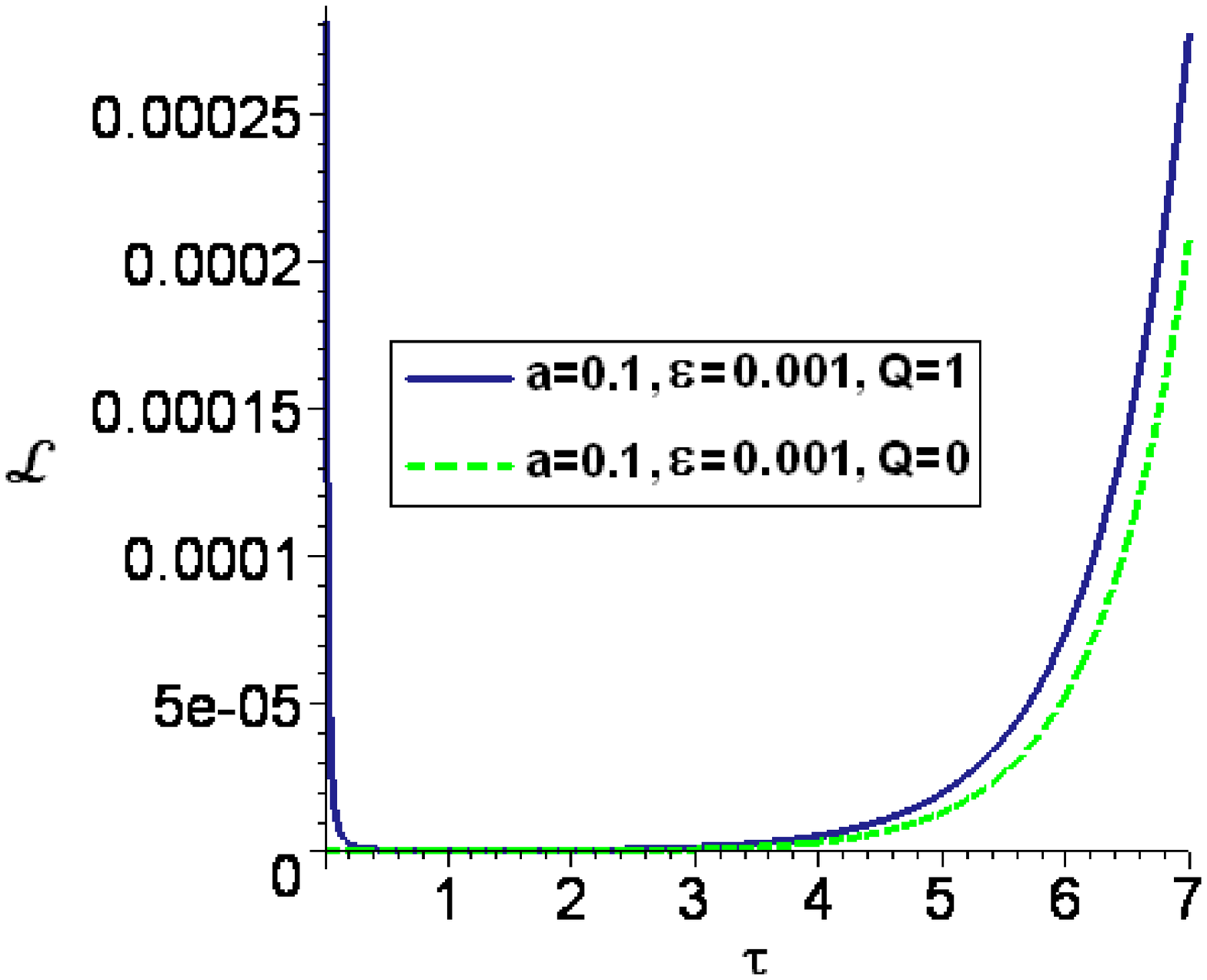}
\hfill
\includegraphics[width=0.45\linewidth]{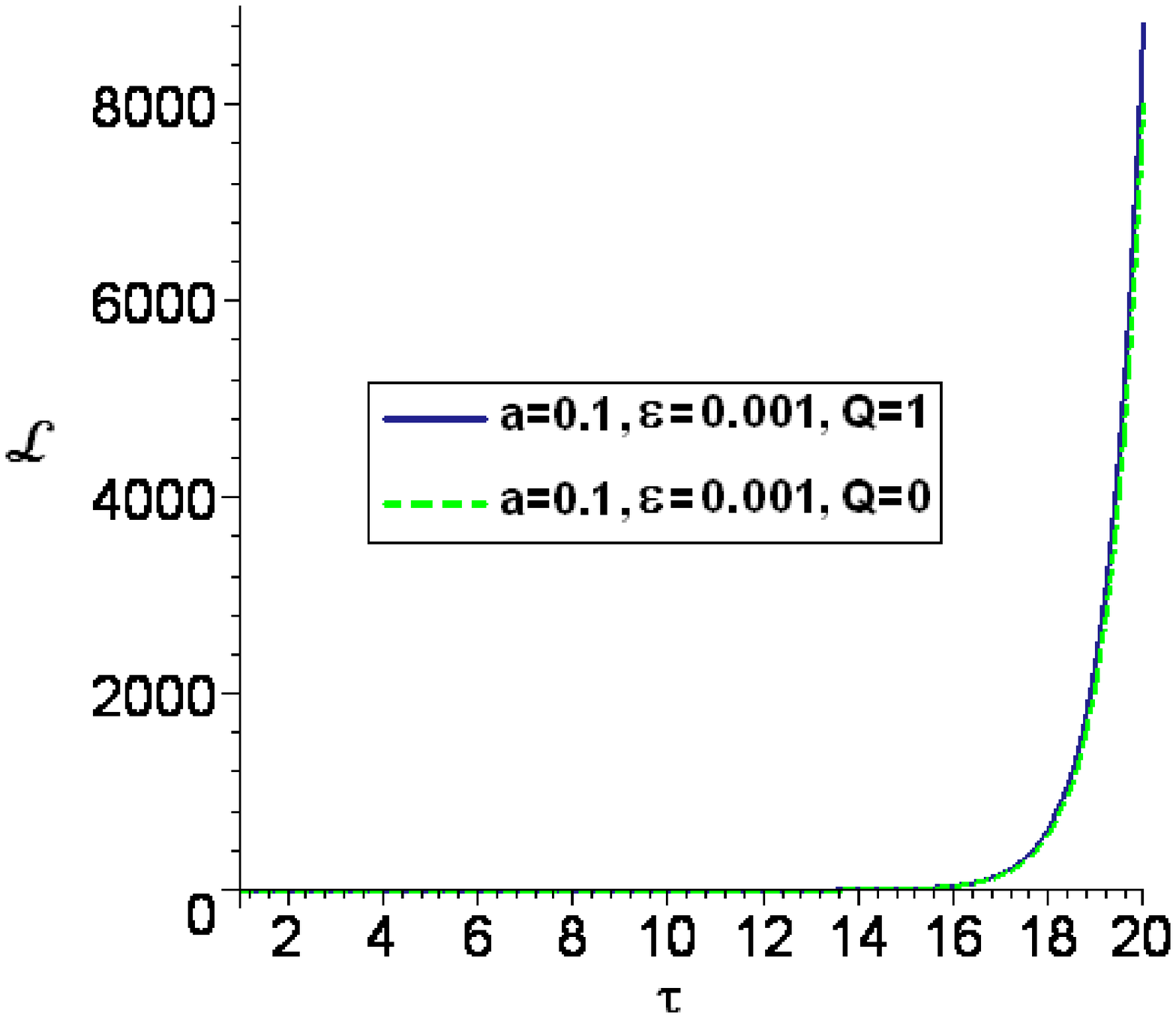}\\
\includegraphics[width=0.45\linewidth]{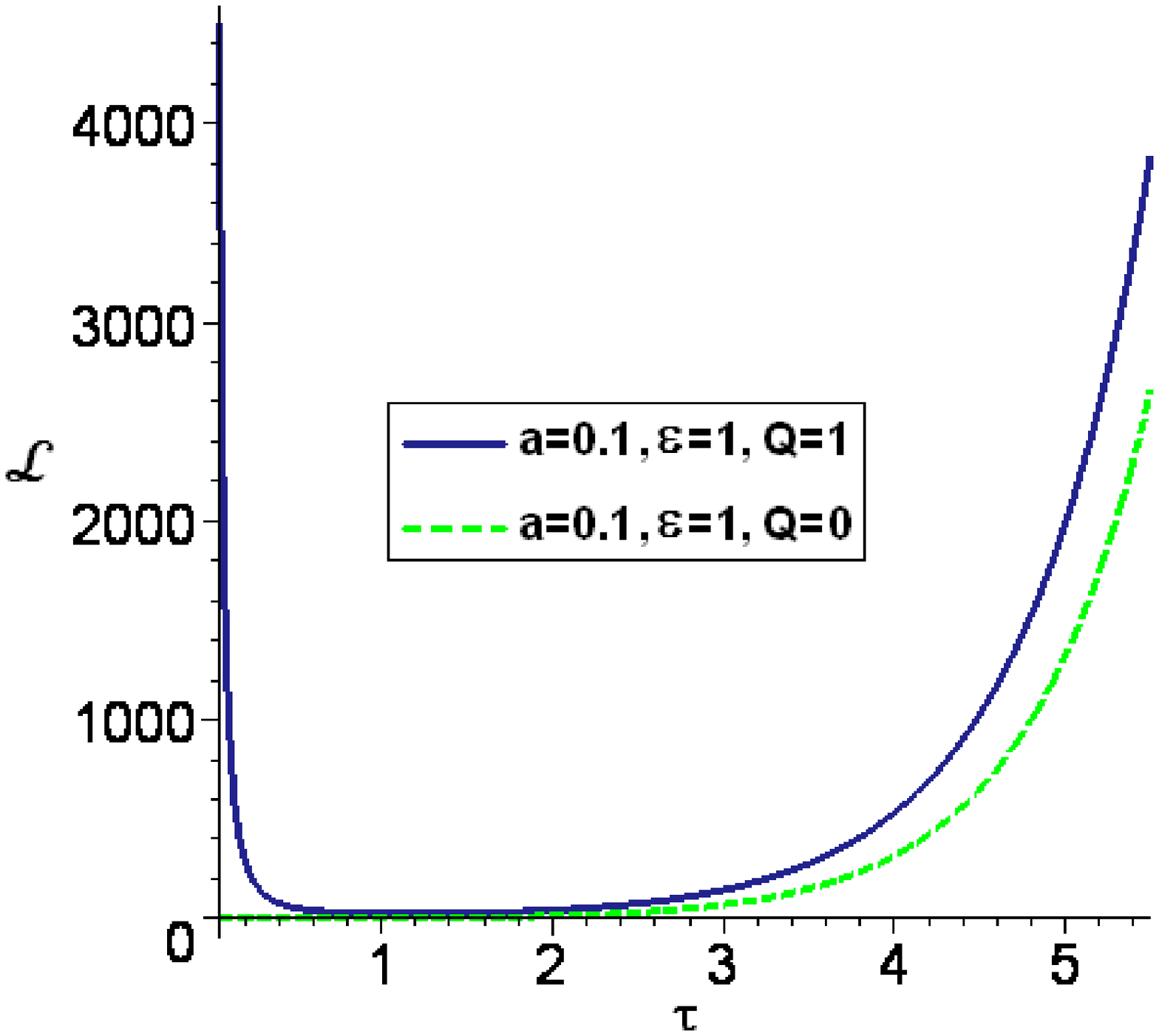} \hfill 
\includegraphics[width=0.45\linewidth]{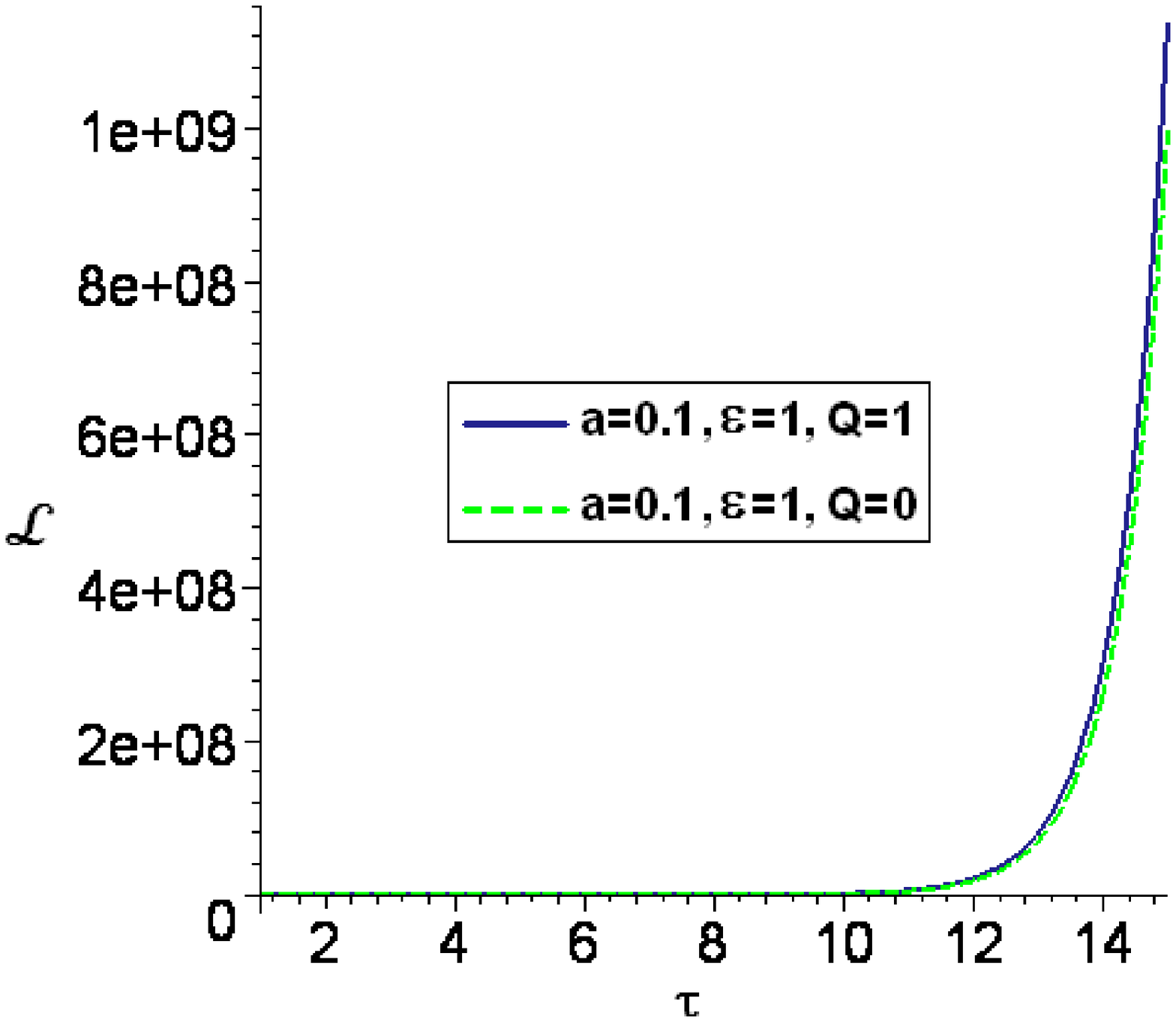}\\
\caption{The KS Lagrangian in the IR and the UV for $\epsilon=0.001$ and $\epsilon=1$}\label{fig:ksactions}
\end{figure}

Let us address a technical issue. 
A problem to deal with here is the singularity in the KS background 
for $Q\neq 0$ (see in the appendix subsection \ref{sssec:Q0} for details). 
Fortunately,
although conceptually an important point, the difference between the 
KS actions with and without $Q$ 
are qualitatively distinguishable only 
in the infrared regime 
(see for instance figures 
\ref{fig:ksactions}). In the UV regime, the leading asymptotic behavior is governed by $P$ and not $Q$.

The IR behavior is easy to understand. In this region, almost at the same value of $\tau$,  
the Lagrangian with $Q=0$ has a zero  
and the corresponding Lagrangian with $Q\neq 0$ has a minimum. 
We therefore switch at this point from the integrand 
with $Q\neq 0$ to the one with $Q=0$
as a regularization of the procedure. 
From the quantitative point of view,  
for $Q=0$
the infrared regime gives a negative contribution to the integral 
which increases slowly with $\epsilon$. 
Thus, for a fixed $a$, 
if for small $\epsilon$ the cascading black hole action dominates 
the difference of the actions, 
this is a reliable result because the negative contribution to the integral 
is negligible. 
On the other hand, if for large $\epsilon$ the KS action dominates, 
that is certainly true, because it does so in spite of the 
negative contribution to the integral. 
 
Finally, the results of the actions comparison are presented 
in fig.\ref{fig:hp}. 
Curves represent the variation with $\epsilon$ of the KS action. 
Each curve with a given color must be compared 
with the horizontal line of the same color, 
i.e, with the cascading black hole with the corresponding temperature. 
As we can see, in each of these cases, 
the deconfined phase dominates for small values of $\epsilon$, 
while the confined phase dominates for large values of this parameter. 
The larger the value of $a$,  
the larger the value of $\epsilon$  
for the phase transition to take place. 
 
A limitation of the approach  described here is that it works for relatively small values of $\epsilon$, 
implying relatively small values of $a$. 
Otherwise, the correction given by (\ref{eq:tau_corr}) might stop to being small.  
Moreover, the difference between the KS solutions 
with $Q=0$ and $Q\neq 0$ could contribute as well. 
Our calculations indicate that the transition can be detected for values as large as $\epsilon=10$  
($a \sim 1$).

\section{Some properties of a confining field  
theory  at finite temperature via gauge/gravity duality}\label{sec:properties}

\subsection{The running of the gauge couplings at finite temperature} 
One of the most interesting properties of the cascading solutions is that they
encode the running of a gauge coupling. 
In particular the matching of the supergravity modes with the two gauge 
couplings takes the form \cite{hko}:
\bea
\frac{4\pi^2}{g_1^2} + \frac{4\pi^2}{g_2^2}&=&\frac{\pi }{g_s e^{\Phi}}, \nonumber \\
\frac{4\pi^2}{g_1^2} - \frac{4\pi^2}{g_2^2}&=&\frac{1}{g_s e^\Phi}
\bigg[\frac{1}{2\pi \alpha'}\left(\int\limits_{S^2}B_2\right) -\pi\bigg].
\eea
For the KT and KS solutions one finds a result that exactly matches the field theory value of the running coupling for 
a gauge theory with $SU(N+M)\times SU(N)$ gauge group. 

\begin{figure}[htp]
\centering
\includegraphics[width=0.4\linewidth]{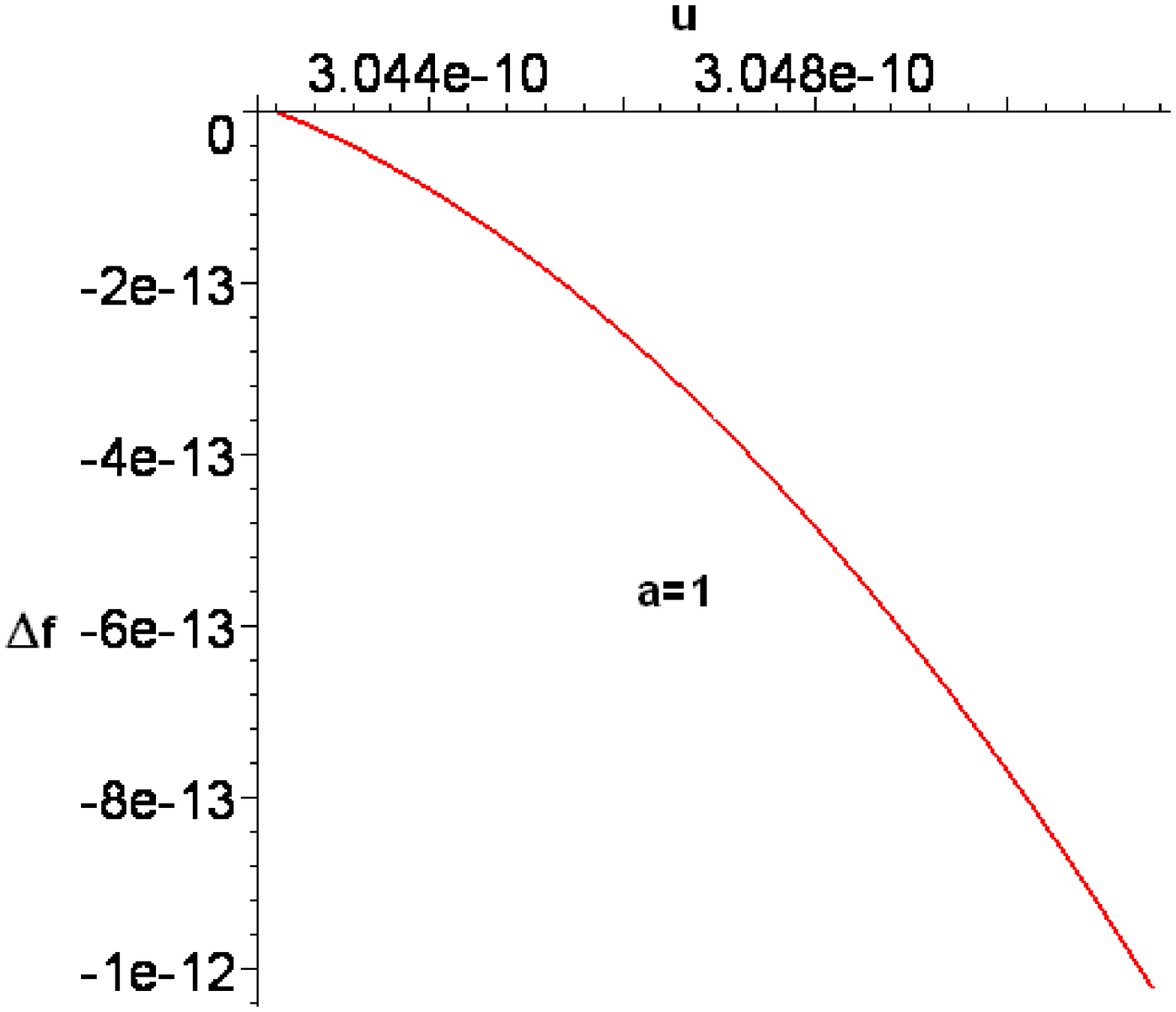}
\hfill
\includegraphics[width=0.4\linewidth]{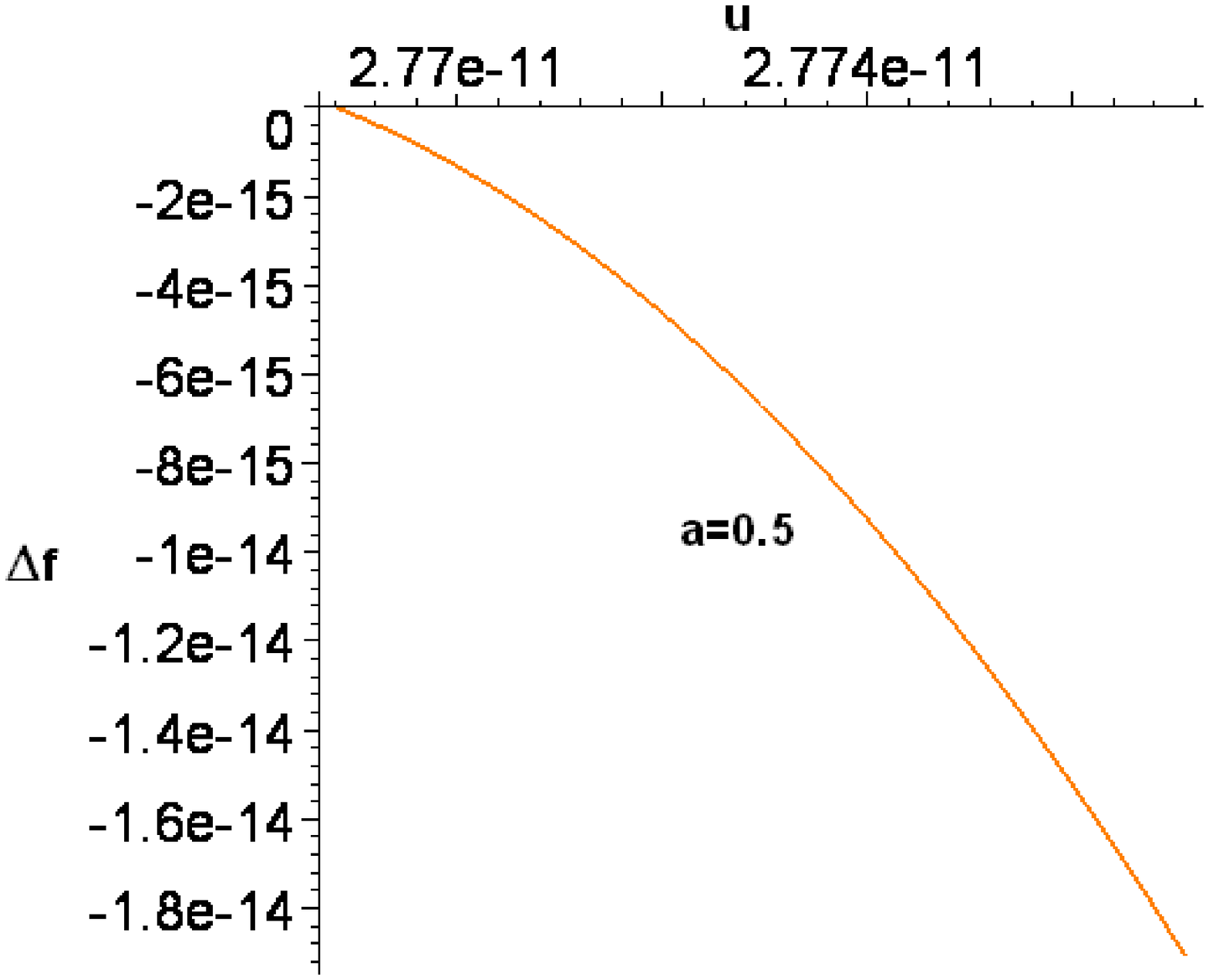}\\
\includegraphics[width=0.4\linewidth]{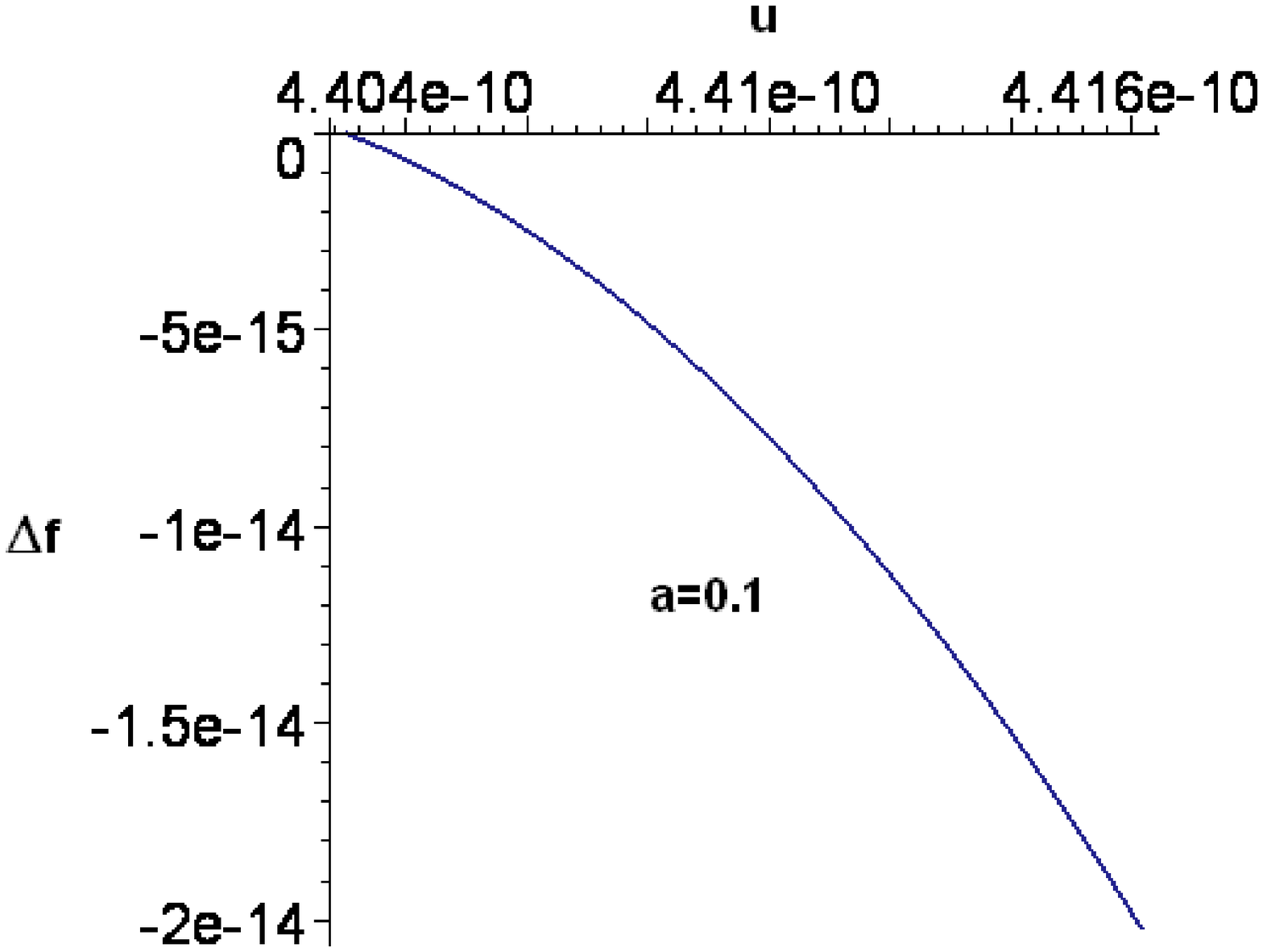}
\hfill 
\includegraphics[width=0.4\linewidth]{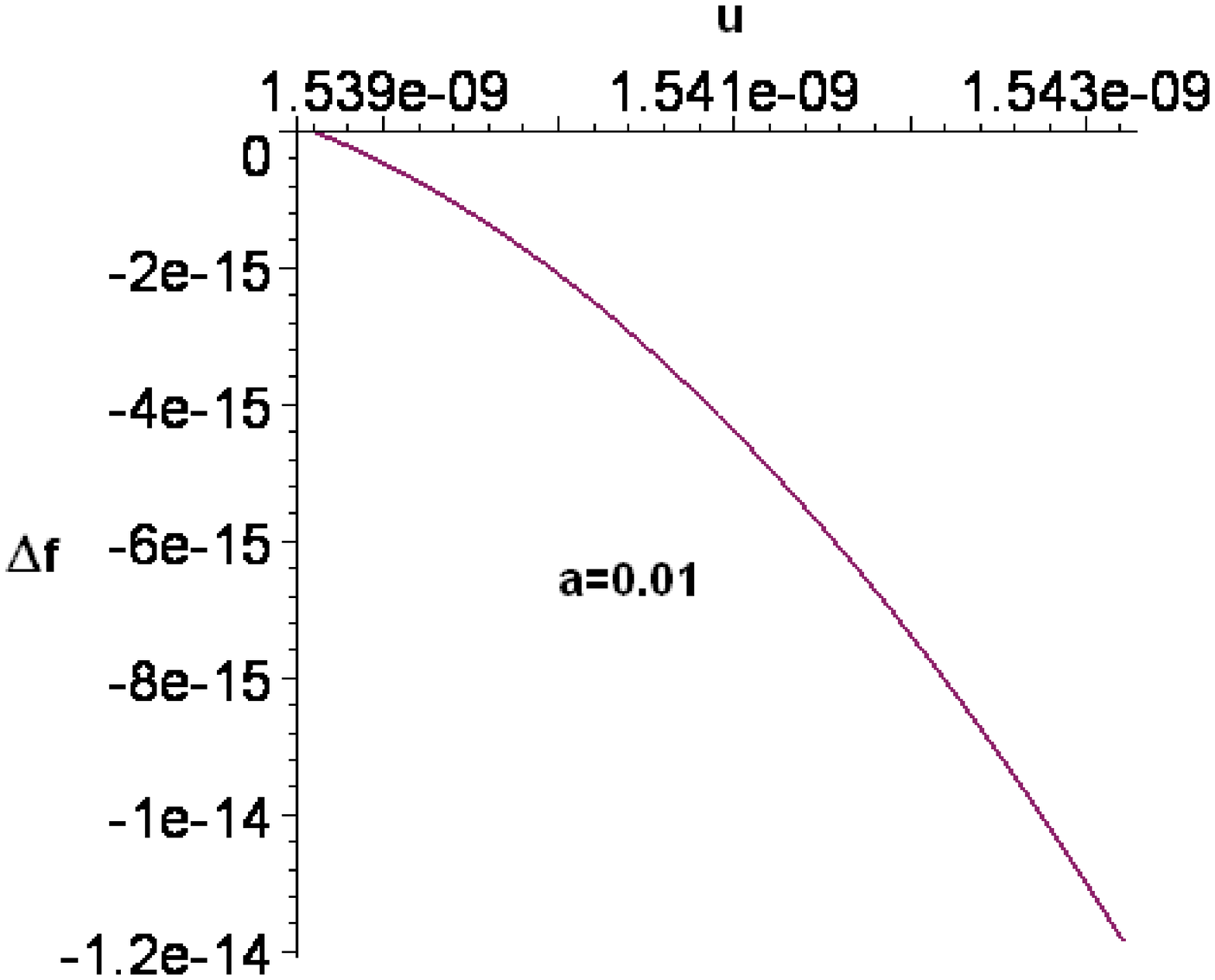}\\
\caption{Behavior of $\Delta f$ for $a=1,\,\,0.5,\,\, 0.1, \,\,0.01$. 
Since it is clearly below our `zero' of $10^{-10}$,
this shows that equation (\ref{eq:run}) represents very accurately 
the form of $f(u)$ in the UV-limit.
}\label{fig:fr}
\end{figure}

For the cascading black hole solutions  we can assume that a similar correspondence holds and evaluate  $B_2$. The 
integral of $B_2$ over the appropriate two-cycle is proportional to the function $f(u)$ in (\ref{eq:forms}). 
Since our solution is numerical we need first to find out 
what is the behavior of $f(u)$ in the UV regime.
If we assume that as $u\rightarrow 0$, 
\begin{equation} 
\label{eq:run} 
f(u) \approx -\frac B4 P {\rm e}^{\Phi(u)}\ln(au) + f_r\, , 
\end{equation} 
where $f_r$ is a constant, 
then criterion (\ref{eq:UVcriterion}) fixes 
\[ 
B = \mathop {\lim }\limits_{u_{sing} \to 0} 4\frac{{\rm e}^{4(y_0+w_0)}}{\Phi_1+\frac 1u_{sing}} \, . 
\] 
The pole in this expression is expected to be cancelled by the logarithmic singularity in Eq.(\ref{eq:run}). 
Substituting $B$ in (\ref{eq:run}), we compare the behavior of the resulting expression with the numerical solution for $f(u)$ near $u_{sing}$. 
Denoting their difference as $\Delta f$, the obtained results are presented in figure (\ref{fig:fr}) for all four cases in table \ref{tab:figpars}. 

It can be seen, 
in the UV regime, that   
up to corrections well below $10^{-10}$, 
$f_r$ is  indeed a constant. 
The approximated values of $f_r$ corresponding to the analyzed cases are
$\{3.22186396, 2.22309019, 0.99140974, 0.41584279\}$.

Thus, the running of the difference of the inverse squares of the coupling constants have several interesting properties. 
First, it is proportional to $P$ which is the parameter breaking the conformal invariance. More importantly, we have the 
following 

\be
\ln(a u)\sim \ln \left( \frac{\rho_0^4}{\rho^4}\right)\sim \ln \left(\frac{T}{\Lambda}\right).
\ee
In the above expression we have used that $2a=\rho_0^4$, where $\rho_0$ is the position of the horizon in the standard 
nonextremal D3-brane solution. Further $\rho_0=T\pi R^2$ for nonextremal D3 branes and finally $u\approx 1/4\rho^4$ in the 
uv region. 

Thus, we conclude that the gravity dual of the cascading black hole has 
\be
\frac{1}{g_1^2}-\frac{1}{g_2^2}\sim P\ln \left(\frac{\Lambda}{T}\right).
\ee

It is worth stressing that the high temperature behavior of the gauge coupling in Yang-Mills theories depends on the 
renormalization schemes at finite $T$. There also seem to be some discrepancies between the imaginary and real time formalism. 
In a sense our result should be interpreted as the string theory prediction and it coincides with the general perturbative 
reasoning \cite{kapusta,bellac}.

\subsection{The viscosity bound} \label{ssec:visc}
 
In this section we simply note that the solution \cite{PT} satisfies 
the condition of \cite{alexuniversality} which implies the viscosity 
bound.  
Namely, we find that for the metric (\ref{metricPT}) the Ricci tensor 
satisfies 
\be 
R^0{}_0-R^1{}_1=4 \, e^{2z-10y}\,\,\,\frac{d^2 x(u)}{du^2}. 
\ee 
This expression is identically zero due to the equation of motion of 
$x$.  
  
In principle, it has been argued that the above expression is sufficient to 
guarantee the viscosity bound \cite{alexuniversality,alexjim}. 
We will nevertheless proceed to check the 
bound explicitly. The main reason being that our background is cascading and that might involve 
subtleties not considered in previous analysis and also that \cite{alexuniversality} used some nonconvariant counterterms. Since our solutions satisfy the criterion (\ref{eq:yzcond}), 
we obtain for the shear diffusion constant, 
\[ 
{\it D}=\frac{|\alpha-3|}{16\pi}\frac 1T \, .
\]  
Assuming the black hole temperature to be given by (\ref{eq:BHT}), 
for the viscosity-entropy ratio,  
\[ 
\frac \eta s = \frac{|\alpha-3|}{16\pi} \, . 
\] 
\begin{figure}[htp]  
\centerline{\includegraphics[width=0.55\linewidth]{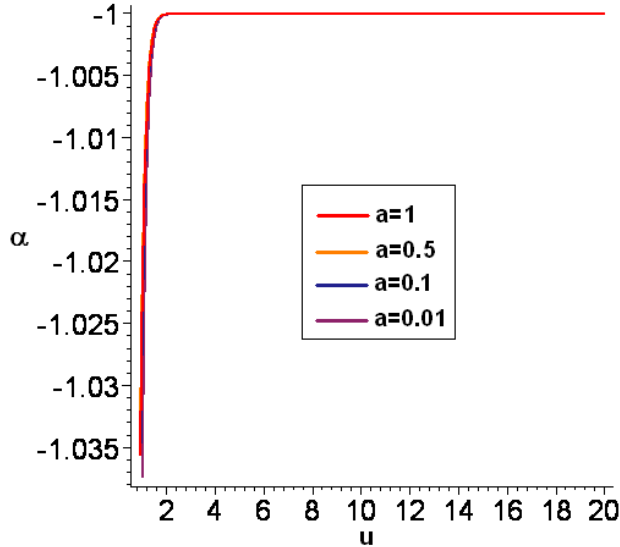}}  
\caption{Behavior of the parameter $\alpha$ as $u\rightarrow u_{hor}$.}   
\label{fig:alpha}  
\end{figure} 
In Fig (\ref{fig:alpha}) we can observe  
that, for all the four cases in table \ref{tab:figpars},
$\alpha\rightarrow - 1$ from below as $u$ approach the horizon.
In this figure the values of $u$ were conveniently rescaled for each case 
using the ratio between the different values of $a$. 
The result seems to indicate that indeed the inequality is saturated. 
We must note here that, in all the cases we analyzed, 
the value to which $\alpha$ converges is actually slightly bigger 
than $-1$.  
For these four cases we obtained  
$\{-0.999999937,-0.999999946,-0.999999977,-0.999999997\}$.  
Nevertheless, since the difference with respect to $-1$ is too close 
to our numerical `zero' 
we regard this effect as a numerical artifact 
due to the proximity to the horizon.

\subsection{Drag force}\label{sec:generaldrag} 

 
In this section we apply the analysis of \cite{seattle,gubser} to discuss the drag force on a quark moving in 
a strongly coupled plasma. We emphasize the general conditions for a supergravity background to admit the motion of a  
classical string that can be interpreted as a drag force on a moving quark.   
Our analysis mirrors the generality discussed in \cite{herzog}, however, here 
we do not assume $AdS$ asymptotics, in particular we aim at understanding cascading backgrounds.  

We consider here a five-dimensional subspace of a supergravity background dual to a field theory. 
\be 
\label{metric} 
ds^2=-G_{00}X_0^2+G_{xx}(dX_1^2+dX_2^2+dX_3^2)+G_{uu}du^2,  
\ee  
where the
metric components $G_{MN}$ are functions of the radial 
coordinate $u$ only. Following \cite{seattle,gubser}, we assume the world-sheet to be embedded as $t=X_0$ 
and $\s=u$ and we allow for   
\be 
X_1=vt+\xi(u), 
\ee  
which means that the end of the string is moving with velocity $v$. In the field theory side we interpret the end of the  
string as a quark moving with velocity $v$.  The Nambu-Goto action is  
\be  
S=\f{1}{2\pi\a 
'}\int dtd\s \sqrt{[G_{00}(G_{uu}+G_{xx}{\xi '}^2)-G_{xx}G_{uu}v^2]}.  
\ee  
Since 
the action does not depend explicitly on $\s$, 
one has that the 
conjugate momenta $\frac{\partial L}{\partial \xi '}$ is a constant, 
where $L$ is the Lagrangian,  
\be  
\Pi _{\xi}= \frac{\partial L}{\partial \xi '}=-\f{G_{00}G_{xx}\xi 
'}{\sqrt{G_{00}(G_{uu}+G_{xx}{\xi '}^2)-G_{uu}G_{xx}v^2}} . 
\ee  
It can be 
rearranged to obtain  
\be  
\xi '=\Pi_{\xi}\sqrt{\f{G_{uu}(G_{00}-G_{xx}v^2)}{G_{00}G_{xx}(G_{00}G_{xx}-\Pi _{\xi}^2)}}.  
\ee 
A key simplifying observation is that, since $\xi '$ cannot be imaginary, we need both expressions in 
numerator and denominator to flip signs simultaneously. This fixes 
$\Pi _{\xi}$ to be  
\be 
\label{Pi}  
\Pi _{\xi}^2=G_{00}(x)G_{xx}(u_*).  
\ee  
Here, 
$u_*$ denotes the radial coordinate satisfying   
\be 
\label{u*} 
G_{00}(u_*)=G_{xx}(u_*)v^2. 
\ee  
The rate of change of momentum is calculated 
to be  
\be  
\f{dp_1}{dt}=\sqrt{-g}{T^{u}}_{x_1},  
\ee  
where
\be 
{T^u}_{x_1}=-\f{1}{2\pi\a '}G_{x_1\nu}g^{u\a }\p _{\a}X^{\nu}.  
\ee 
Here, $G_{ij}$ denotes the metric in (\ref{metric}), while 
$g_{ij}$ denotes the induced metric on the world-sheet. After some 
algebraic simplifications, we obtain  
\be 
\label{DF1} 
\f{dp_1}{dt}=-\f{1}{2\pi\a '}\Pi _{\xi}.  
\ee  
We can rewrite 
of above equation as  
\be 
\label{DF2}  
\f{dp_1}{dt}=-kp_1, 
\ee  
if 
condition (\ref{u*}) dictates  $\Pi _{\xi}$ to be proportional to 
$p_1$. In the above equation, $k$ is the drag force constant.

\subsubsection*{Conditions for drag force} 
 
Let us discuss the conditions under which we can observe a drag force  
proportional to the momentum.  
We need to solve for a radial position  $u_*$  such that  
\be 
v^2=G_{00}(u_*)/G_{xx}(u_*). 
\ee 
The expression for the canonical momentum is  
\be 
\Pi_\xi = v G_{xx}(u_*(v)). 
\ee 
So, the question for a drag force essentially boils down to  
verifying that  
\be 
\label{criterion} 
G_{xx}(u_*(v))=\frac{b}{\sqrt{1-v^2}}, 
\ee 
where $b$ is a constant.

Two comments are in order about the relevant energy scales and the speed of the quarks:

\noindent 
$\bullet$ The nonrelativistic motion of the probe corresponding to 
small $v$ is localized near the points where  
$G_{00}$ vanishes. This corresponds, generically, to the horizon.  
In terms of energy scales,  
it corresponds to the infrared region.

\noindent 
$\bullet$ Relativistic velocities $v$ close to the speed of light 
correspond  
to radial positions for which  
$G_{00}\approx G_{xx}$. Generically, this is the asymptotic region, which in terms of the field theory is the ultraviolet region and corresponds to the near conformal limit.

\subsubsection*{Non-extremal D3 brane} 
 
In the case of non-extremal D3 branes we have that  
\be 
G_{00}=u^2 \left(1-\frac{u_0^4}{u^4}\right), \qquad G_{xx}=u^2. 
\ee 
The equation that determines the position $u_*$ as a function of the velocity of the end  
of the string is thus: 
\be 
v^2=\frac{G_{00}}{G_{xx}}=\left(1-\frac{u_0^4}{u_*^4}\right). 
\ee 
Thus, we find  
\be 
u_*^2=\frac{u_0^2}{\sqrt{1-v^2}}. 
\ee 
We can now easily verify that  
\be 
G_{xx}(u_*(v))=u_*^2=\frac{u_0^2}{\sqrt{1-v^2}}. 
\ee 
This expression coincides with the criterion (\ref{criterion}). This case was discussed in Refs.\cite{seattle, gubser}.

\subsubsection*{Cascading black holes}
\label{sssec:bhdf} 
 
Here we simply adjust the computation of the drag force 
presented,  
for example, in \cite{seattle,gubser}. The calculations presented here were, in particular, performed in 
\cite{alexjet,elenajet} without using the full metric of the cascading black hole.  
The AdS part of the metric in consideration in string frame is  
\be 
ds^2=e^{\P 
/2}[e^{2z}(-e^{-6x}dX_0^2+e^{2x}dX_idX^i)+e^{10y-2z}du^2]  
\ee  
We consider the worldsheet to be along $t=X_0$ and $\s=u$ directions.  
Following the general arguments at the beginning of this subsection 
\be\label{eq:Pi}  
\Pi_{\xi}=\sqrt{v}e^{\frac{\P}{2}+2z}(u(x(v)))  
\ee  
Here, $u$ is chosen to satisfy 
$v=e^{-4x(u)}$ for a given $v$. 
This relation is to be viewed as an equation for the  
radial coordinate $u$, it defines a particular value $u_p$. 
We need to 
  evaluate relation (\ref{eq:Pi}) at $u=u_p$ and, in turn, 
find $\Pi _{\xi}$ as a function of $v$.
Further, to establish the  
existence of a drag force we need to verify the existence of a $b$ such that   
\be 
\label{eq:AnzPi} 
\Pi_{\xi}= b \frac v{\sqrt{(1-v^2)}}, 
\ee  
which thus implies that $\Pi_{\xi}=\f{b}{m}p_1$ 
and directly leads to a drag force parameter 
using equations (\ref{DF1}) and (\ref{DF2}). 
With this aim, 
equating equations (\ref{eq:Pi}) and (\ref{eq:AnzPi}) we obtain, 
\bea 
\label{eq:b} 
b(u) &=& \sqrt {2}{{\rm e}^{\f\Phi 2 +2\,z}}
\sqrt {\sinh \left( 4\,au \right) }\, \nonumber\\ 
v(u) &=& {\rm e}^{-4au}\, . 
\eea 
First, note that in the IR regime ($u\rightarrow\infty$), 
\be 
b \rightarrow {{\rm e}^{\f{\Phi(u)}2}}
{{\rm e}^{2\,{\it z_*}}}{{\rm e}^{2\, \left( \alpha+1 \right) a u}}\, , 
\ee 
where we used the asymptotical expression (\ref{eq:yzasym}) for $z(u)$. 
In the pure D3 case, 
where $\Phi(u)=0$ and $\alpha=-1$, 
 we have that $b={\exp(2\,{\it z_*}})$. 
Recalling that, 
as shown in subsection \ref{ssec:visc},
for our solution $\alpha\approx -1$,
and noting that the dilaton is very weak near the horizon
(for instance, for our four cases we obtained 
$\Phi_* \approx \{-3\times 10^{-7},-2\times 10^{-7},
-1.2\times 10^{-7},-1.4\times 10^{-8}\}$), 
we see that for the cascading black hole $b \approx$ constant in the IR-limit. 

On the other hand, knowing that as $u\rightarrow 0$, 
$z(u)$ behaves as described by Eq.(\ref{eq:zUV}), we obtain, 
\be 
b \rightarrow \sqrt{2a} {\rm e}^{\f {\Phi_0}2}\, , 
\ee 
implying that $b$ is finite in the UV-limit. 
 
\begin{figure}  
\centerline{\includegraphics[width=0.55\linewidth]{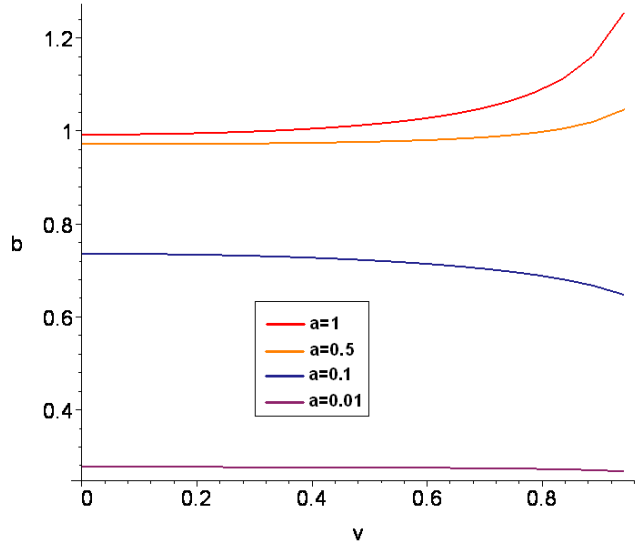}}  
\caption{Behavior of the drag force parameter $b$ as a function of the quark velocity $v$.}   
\label{fig:df}  
\end{figure}

This way,  
we have shown {\it analytically} that when the string explores any of the two  
asymptotic regions, we verify the existence of a drag force.  
For a given temperature
the variability of $b$ from one regime to the other is mainly related 
to the dilaton variability.  
In fig.\ref{fig:df}  
we show the drag force coefficient $b$ as given by Eqs.(\ref{eq:b})
for all values of $v$,
that is, for all velocities of the external quarks. 
The curves represent the four cases in table \ref{tab:figpars}. 
It is worth stressing that in the 
relativistic regime there is a clear velocity dependence in the coefficient $b$. However, 
for the range of values of temperatures that we have explored it, it does not seem to be 
universal. In particular, the fig.\ref{fig:df} shows that in the regime of relativistic 
velocities some curves have inflections upwards while other bend downward. 
Clearly, it seems to depend on the temperature and deserves further study.

\section{Discussion and outlook}\label{sec:conclusions} 
 
In this paper we have established numerically the existence of a 
Hawking-Page phase transition between the supergravity solution of 
\cite{PT} which represents a black hole in a cascading background and the Klebanov-Strassler solution.   

We also studied some properties of the deconfined phase given by our cascading black hole solution. For instance, we have found that the difference of the inverse squares of the coupling constants is proportional to $P$ (which is the parameter breaking the conformal invariance) and that it runs as $\ln(\Lambda/T)$.

We also verify that our solution seems to satisfy the well-known bound for the viscosity to entropy ratio.  
 
Finally, we have explicitly shown that the universal form of the horizon in 
this theory  guarantees the existence of a drag force for a 
fundamental string moving in this background and in the  dual field 
theory that can be interpreted as jet quenching. The near horizon region describes 
nonrelativistic motion of the quarks and since the near horizon geometry has been shown to be 
universal we conclude that the nonrelativistic drag force on the quarks is also universal in the 
class of cascading theories. In the relativistic region we also found a well-defined drag force parameter. 
Numerically, we observed a velocity dependence for this parameter in the relativistic region. 
 
It is important to stress that most of the calculations performed in this paper 
 requires knowing the 
solution for all values of the radius and  that an analysis solely 
based on the asymptotic near the horizon is not reliable. 
 
Let us conclude with listing a number of possible directions. 
Cascading theories, along the lines of the particular model 
described here are generic in the AdS/CFT. In fact,  we believe that 
there might be infinite classes of them based on some results 
presented in \cite{ehk,ypqcas}. The  fact that the cone over 
$Y^{p,q}$ does not admit complex deformations seems like a deterrent 
for a construction  directly mimicking the warped deformed  conifold 
of Klebanov and Strassler but a more  general solution where the 
fluxes play  a crucial role are not ruled out. 
Turning on temperature of these solutions should not 
change substantially the asymptotic UV form obtained in \cite{ehk,ypqcas}.
 
It would be interesting to better understand the charges in the  
comparison between the cascading black hole with the KS 
background. Basically the KS solution has $Q=0$, that is, the number of 
D3 branes at the tip deformed conifold equal to zero. As shown in \cite{ks} and reviewed in 
section \ref{sssec:Q0}, we see that nonzero $Q$ results in a 
singularity at the origin. Another interesting direction would be the construction of black holes in the backgrounds with back 
reacted flavor such like the one described in Ref.~\cite{backreactedflavor}. 
In particular, some of these backgrounds should be natural 
generalizations of the construction of \cite{PT}.

While this manuscript was in the final stage of its preparation, a new paper was 
posted where the authors claim to have found evidence for a transition in a solution 
with a black hole in the IR and KT in the UV \cite{alexofer}. We plan an in-depth 
discussion of the relation between our and their solutions. 
Preliminarily, let us note here 
that there are some key differences between our solutions. 
First, 
as far as we understand, to find their solution they need to numerically 
look for a zero of a function in ten variables, i.e., the norm of their `mismatch vector'. 
Since this is a positive semi-definite function, assuming that this zero exists, this problem 
can be translated into finding the global minimum of a 10D-surface. 
Nevertheless, this is a case 
of constrained optimization, since the Hamiltonian constraint is also valid in ten 
dimensions, 
where it shows as a reparametrization of the radial coordinate. 
However, 
the authors 
of \cite{alexofer}  made no mention to whether their numerical solution satisfies this constraint. As 
opposed to that, our solutions are forced to automatically satisfy the zero-energy 
constraint and that this certainly happen is one of the main criteria of quality we 
use to certify our numerical output. 
This was first explained in Ref. \cite{PT} and emphasized in this manuscript. Secondly, 
their solution seems to have exactly KT in the boundary (we note here that the KT solution 
does not satisfy the Hamiltonian constraint for $a \neq 0$) and to erase the large $u$ 
singularity typical of KT (we called this singular point as $u_{sing}$), pasting the black hole metric in the IR 
solution somewhere in the 'middle' of the radial domain. Our solution extents from $u_{sing}$ to 
infinity and, as explained in Section \ref{sec:completeness}, we set $u_{sing}=0$ to guarantee the 
completeness of our background.

\section*{Acknowledgments} 
 
We thank Alex Buchel for various comments, Elena C\'aceres for comments on the drag force 
and Javier Mas for various useful comments and explanations about related matters.  
This work is  partially supported by Department of Energy under  
grant DE-FG02-95ER40899 to the University of Michigan  
and by grant CNPq/CLAF-150548/2004-4. C. T-E thanks the MCTP for 
hospitality during the final stages of this work and  LPZ thanks Universidad de Santiago de Compostela for 
a very warm hospitality.

\appendix

\section{The Klebanov-Strassler background}

The KS metric is of the form: 
\bea  
\label{specans} 
ds^2_{10} &=&   h^{-1/2}(\tau)   dx_n dx_n 
 +  h^{1/2}(\tau) ds_6^2 \ , \nonumber \\ 
ds_6^2 &=& \frac 12\varepsilon^{4/3} K(\tau) 
\Bigg[ \frac 1{3 K^3(\tau)} (d\tau^2 + (g^5)^2)\nonumber \\ 
 &+& 
\cosh^2 \left(\frac{\tau}{2}\right) [(g^3)^2 + (g^4)^2] 
+ \sinh^2 \left(\frac{\tau}{2}\right)  [(g^1)^2 + (g^2)^2] \Bigg] 
\ , 
\eea 
the one-forms $g^i$ are defined as  
 
\bea \label{fbasis} 
g^1 = \frac{e^1-e^3}{\sqrt 2}\ ,\qquad 
g^2 = \frac{e^2-e^4}{\sqrt 2}\ , \nonumber \\ 
g^3 = \frac{e^1+e^3}{\sqrt 2}\ ,\qquad 
g^4 = \frac{e^2+ e^4}{\sqrt 2}\ , \nonumber \\ 
g^5 = e^5\ , 
\eea 
where 
\begin{eqnarray} 
e^1\equiv - \sin\theta_1 d\phi_1 \ ,\qquad 
e^2\equiv d\theta_1\ , \nonumber \\ 
e^3\equiv \cos\psi\sin\theta_2 d\phi_2-\sin\psi d\theta_2\ , \nonumber\\ 
e^4\equiv \sin\psi\sin\theta_2 d\phi_2+\cos\psi d\theta_2\ , \nonumber \\ 
e^5\equiv d\psi + \cos\theta_1 d\phi_1+ \cos\theta_2 d\phi_2 \ . 
\end{eqnarray} 
 
The function in the metric is 
\be 
{\cal K}(\tau)= \frac{ (\sinh (2\tau) - 2\tau)^{1/3}}{2^{1/3} \sinh \tau} 
\ . 
\ee 
For asymptotically large values of $\tau$ one can introduce a natural radial  coordinate  
$r^2 = \frac{3}{2^{5/3}} \varepsilon^{4/3} e^{2\tau/3}$ such that the metric on the  
conifold becomes  
$ ds_6^2 \approx dr^2 + r^2 ds^2_{T^{1,1}}$.
 
The matter fields are as follow: 
\begin{eqnarray} 
F_3 &=& \frac{M\alpha'}{2} \left \{g^5\wedge g^3\wedge g^4 + d [ F(\tau) 
(g^1\wedge g^3 + g^2\wedge g^4)]\right \} \nonumber \\ 
&=& \frac{M\alpha'}{2} \left \{g^5\wedge g^3\wedge g^4 (1- F) 
+ g^5\wedge g^1\wedge g^2 F \right. \nonumber \\ 
&& \qquad \qquad \left. + F' d\tau\wedge 
(g^1\wedge g^3 + g^2\wedge g^4) \right \}\ , 
\end{eqnarray} 
with $F(0) = 0$ and $F(\infty)=1/2$, and 
\be 
B_2 = \frac{g_s M \alpha'}{2} [f(\tau) g^1\wedge g^2 
+  k(\tau) g^3\wedge g^4 ]\ , 
\ee 
\begin{eqnarray} 
H_3 = dB_2 &=& \frac{g_s M \alpha'}{2} \bigg[ 
d\tau\wedge (f' g^1\wedge g^2 
+  k' g^3\wedge g^4)   
\nonumber \\ 
&& \left. + \frac 12 (k-f) 
g^5\wedge (g^1\wedge g^3 + g^2\wedge g^4) \right]\ . 
\end{eqnarray} 
 
The self-dual 5-form field strength may be 
decomposed as $\tilde F_5 = {\cal F}_5 + \star {\cal F}_5$. We 
have 
\be 
{\cal F}_5 = B_2\wedge F_3 = \frac{g_s M^2 (\alpha')^2}{4} \ell(\tau) 
g^1\wedge g^2\wedge g^3\wedge g^4\wedge g^5\ , 
\ee 
where 
\be 
\ell = f(1-F) + k F\ , 
\ee 
and 
\be 
\star {\cal F}_5 = 4 g_s M^2 (\alpha')^2 \varepsilon^{-8/3} 
dx^0\wedge dx^1\wedge dx^2\wedge dx^3 
\wedge d\tau \frac{\ell(\tau)}{K^2 h^2 \sinh^2 (\tau)}\ . 
\ee

\be 
\alpha =4 (g_s M \alpha')^2 
\varepsilon^{-8/3}\ . 
\ee 

The solution for the functions defining the matter content  is: 
\bea 
F(\tau) &=& \frac{\sinh \tau -\tau}{2\sinh\tau}\ , 
\nonumber \\ 
f(\tau) &=& \frac{\tau\coth\tau - 1}{2\sinh\tau}(\cosh\tau-1) \ , 
\nonumber \\ 
k(\tau) &=& \frac{\tau\coth\tau - 1}{2\sinh\tau}(\cosh\tau+1) 
\ , \nonumber \\ 
\ell(\tau) &=& f(1-F) + kF =  \frac{\tau\coth\tau - 1}{4\sinh^2\tau} 
(\sinh 2\tau-2\tau) 
\ .\eea 
The warp factor with the condition that it vanishes at infinity is  
\be \label{intsol} 
h(\tau) = \alpha \frac{ 2^{2/3}}{4} I(\tau) = 
(g_s M\alpha')^2 2^{2/3} \varepsilon^{-8/3} I(\tau)\ , 
\ee 
where 
\be 
I(\tau) \equiv 
\int_\tau^\infty d x \frac{x\coth x-1}{\sinh^2 x} (\sinh (2x) - 2x)^{1/3} 
\ . 
\label{eq:KSI}
\ee 
Two important limits are: 
\be 
I(\tau\to 0) \to a_0 + O(\tau^2) \ ;  
\ee 
\be 
\ I(\tau\to\infty)\to 3 \cdot 2^{-1/3} 
\left (\tau - \frac 14 \right ) e^{-4\tau/3} 
\ ,\ee 
where $a_0\approx 0.71805$.  At large 
$\tau$ the integrand becomes  
\be 
h\approx  3^4 2^{-4}\frac{(g_sM\alpha')^2}{r^4}\ln  
\left(\frac{2^{5/3}}{3}\frac{r^2}{\epsilon^{4/3}}\right),  
\label{eq:KSIuv}
\ee 
where we have used that for large radius  
$r^2 = \frac{3}{2^{5/3}} \varepsilon^{4/3} e^{2\tau/3}$:

\subsection{The action for the KS model} 
 
A convenient way to look at this solution is through the prism of 
a one-dimensional system as discussed in \cite{resolved}.  
Motivated by the form  of the deformed conifold metric we  
make  the following  ansatz for the metric 
  
\be\la{dddf} 
ds^2 =  e^{2 p-x} (e^{2A}  dx^\mu dx^\mu + du^2)  
  +  
\left[ e^{-6p - x} \gp^2  +   e^{x+y}    ( \go^2 + \gd^2) 
+  e^{x-y}  ( \gt^2 + \gc^2) 
 \right] \ .  \ee 
The ansatz for the p-forms is: 
\be 
H_3  = du\wedge[  f'(u) \go \we \gd + k'(u) \gt \we \gc] \ , \ \ \ 
\la{he} \ee 
\be  
F_3 = F(u) \go\we\gd\we\gp + [2P-F(u)] \gt\we\gc\we\gp + 
F'(u) du \we ( \go \we \gt +  \gd \we \gc) 
\ , \la{hhe} 
\ee 
\be 
 F_5 = \F_5 + \F_5^* \ , \ \ \ \ \ \ \ 
 \F_5 =    K(u) \go\we\gd\we\gt \we\gc\we\gp\ , \la{poi} 
 \ee 
 \be  
 K(u)  \equiv Q +  k(u)  F(u) + f(u) [2P-F(u)] \ , \label{uui} 
 \ee 
 where $F,f,k$ are functions to be determined and $P$ and $Q$ are 
 constants. We  
explicitly ensure that  
the Bianchi identities for the p-forms are satisfied automatically.  
The  1-d action reproducing 
the resulting equations of motion restricted to the above ansatz 
has the following general structure 
\begin{equation} 
 S 
= c  \int du 
\ e^{4 A} \bigg[ 3 A'^2 
- \frac12 G_{ab}(\varphi)  \varphi'^a  \varphi'^b - 
 V(\varphi)\bigg] 
 \ ,  \la{vvv}  \ee 
 where $c = -4  \frac{Vol_9}{2\kappa_{10}^2}$. 
 It should be supplemented with   the ``zero-energy" constraint 
 \be 
 3 A'^2 
- \frac12 G_{ab}(\varphi)  \varphi'^a  \varphi'^b + 
 V(\varphi)= 0 \ . \la{ret} 
 \ee

The action \rf{vvv} is thus a classical mechanical action for the fields  $A$ and  
$\varphi^a=(x,y,p,\Phi,f,k,F)$.  
The corresponding  kinetic and potential terms  
in \rf{vvv} are found to be: 
\be 
G_{ab}(\varphi)  \varphi'^a  \varphi'^b 
= x'^2+ \frac 12 y'^2+6p'^2 +  \frac 14 \bigg[ 
\Phi'^2  +   e^{-\Phi- 2x} 
 ( e^{ - 2y}  f'^2 + e^{ 2y} k'^2) + 2  e^{\Phi- 2x} F'^2  \bigg] \ ,  
 \la{kine} 
 \ee 
 $$ 
V(\varphi) = \frac 14 e^{-4p-4x}- e^{2p-2x}\cosh y  
 + \frac 14 e^{8p}\sinh^2 y  $$ \be  
 +\  
 {\tx \frac 18} e^{8p}  \bigg[ \ha  e^{-\Phi-2x} (f-k)^2  
 +  e^{\Phi-2x}  [ e^{  -2y}  F^2  + e^{ 2y} (2P-F)^2] 
+  e^{-4x} K^2  \bigg]\ ,  \la{pote} 
\ee 
where $K$ is the combination of the independent functions 
$f,k,F$  given in   \rf{uui}. 
  
The first order equations for the independent functions   
 $A,x,y,p, f,k,F,\P$  are: 
 \be 
x'= - e^{-2p-2x}    - \ha  e^{4p - 2x} K  \ , \ \ \ \ \ \ \ \ \ \ 
y' = e^{4p  } \sinh y\ , 
\la{syd} \ee 
\be\la{syyd} 
p'=  {\tx \frac 13} 
e^{4p}\cosh  y  - \frac 16  e^{-2p-2x}  + \frac 16 e^{4p-2x} K \ , \ 
\ee 
\be\la{syyyd}  
A'=-{\tx \frac 13} 
 e^{4p}\cosh y  - {\tx \frac 13} e^{-2p-2x}  
   -   \frac 16   e^{4p - 2x } K \ , 
\ee 
\be 
 f' =    e^{\Phi + 4p +2y } (2P-F)  \ , \ \ \ \ 
k' =   e^{\Phi + 4p - 2y }F \ ,\ \ \   
F' = - \frac12 e^{-\Phi + 4p } (f-k)\ ,   \ \ \ \ \P'=0 \ .  \la{sysd} 
\ee

The functions that we introduced in this subsection are explicitly given in terms of the solution of the 
previous subsection as:  

\[ 
{\rm e}^x = \frac 14 \epsilon^{4/3} {\mathcal K} \sinh(\tau) h^{1/2} ,
\] 
\[ 
{\rm e}^{2p} = 24^{1/3} h^{-1/3} 
\epsilon^{-8/9}{\mathcal K}^{1/3}\sinh(\tau)^{-1/3}, 
\] 
\[ 
{\rm e}^y = \tanh(\frac \tau 2). 
\] 

We will also use explicit forms for the derivative with respect to $u$ 

\bea
A'&=& -\frac16 ({\rm e}^{4p}({\rm e}^y + {\rm e}^{-y})  
+ 2{\rm e}^{-2p} {\rm e}^{-2x} + K{\rm e}^{4p} {\rm e}^{-2x} ), \nonumber \\
p'&=& \frac16 ({\rm e}^{4p}({\rm e}^y + {\rm e}^{-y})  
- {\rm e}^{-2p} {\rm e}^{-2x} + K{\rm e}^{4p} {\rm e}^{-2x} ), \nonumber \\
y'&=& {\rm e}^{4p} \sinh(\ln({\rm e}^{y})), \quad x'= - {\rm e}^{-2p} {\rm e}^{-2x} - \frac K2 {\rm e}^{4p} {\rm e}^{-2x} 
, \qquad \Phi=\Phi'=0,  \nonumber \\
k'&=& F{\rm e}^{\Phi}{\rm e}^{4p}{\rm e}^{-2y}, \qquad  
f'= (2P-F){\rm e}^{\Phi}{\rm e}^{4p}{\rm e}^{2y}, \qquad   
F'= -\frac{f-k}2 {\rm e}^{-\Phi}{\rm e}^{4p}. \nonumber 
\eea 
The expression for the action we need to evaluate is finally of the form: 

\be
{\mathcal L}= -\frac 4h {\rm e}^{2x}{\rm e}^{-4p} (3A'^2-\frac 12 G_{ab} - V), \qquad 
S_{KS}=\int d\tau {\mathcal L}.
\ee

\subsubsection{The KS background has no regular D3 branes at the apex} 
\label{sssec:Q0}
 
As noted in \cite{ks}, the KS-like Ansatz allows for 
general solutions having $K(u)$ with arbitrary $Q$ in (\ref{uui}). In the 
more standard literature $Q$ is identified as $N$. 
These 
solutions are, however, singular. Namely, near the apex of the deformed 
conifold (small values of $\tau$) the warp factor takes the form: 
 
\be 
h\approx \frac{Q}{\tau}. 
\ee 
An alternative way of understanding this singularity is by realizing 
that it corresponds to the freedom to add a homogeneous solution to 
the warp factor. Namely, a solution of the form 
\be 
\frac{1}{\sqrt{g_6}}\partial_\tau 
\left(\sqrt{g_6}g^{\tau\tau}\partial_\tau\tilde{ h}\right)=0,  
\ee 
or  
\be 
\tilde{h}=Q\int\frac{d\tau}{{\mathcal K}^2\sinh^2\tau}. 
\ee 
This last statement allows to interpret the elimination of the 
singularity as a statement about regular  D3 branes. It is possible 
that, as in the interpretation for the resolved conifold of 
\cite{resolved}, this divergence signals a smearing of D3 branes along 
a three-dimensional subspace and therefore could also be cure, not 
just by turning off the charge $Q$, 
but also by localizing the branes 
appropriately as it was done recently in Ref.\cite{smearing}. 
 
This situation presents a conceptual problem. It is natural to compare  
a field theory with parameters $(N,M, \epsilon\sim \Lambda_{strong})$  
with a theory with $(N,M,a\sim Temperature)$. In realistic cases, the scale $\Lambda_{strong}$ is dynamically 
generated but supergravity methods are still far from achieving that. 
However, for KS we have $N=0$ and for the cascading black hole we have $u_{sing}=0$. The  
comparison is really not completely clear. We deal with this point while analyzing the transition 
in the next section.



\begin{thebibliography}{99}   
 
\bibitem{malda} 
  J.~M.~Maldacena, 
 ``The large N limit of superconformal field theories and supergravity,'' 
  Adv.\ Theor.\ Math.\ Phys.\  {\bf 2} (1998) 231 
  [Int.\ J.\ Theor.\ Phys.\  {\bf 38} (1999) 1113] 
  [arXiv:hep-th/9711200]. 
 
 
\bibitem{agmoo} 
  O.~Aharony, S.~S.~Gubser, J.~M.~Maldacena, H.~Ooguri and Y.~Oz, 
 ``Large N field theories, string theory and gravity,'' 
  Phys.\ Rept.\  {\bf 323} (2000) 183 
  [arXiv:hep-th/9905111]. 
 
 
 
\bibitem{igortemp} 
  S.~S.~Gubser, I.~R.~Klebanov and A.~W.~Peet, 
``Entropy and Temperature of Black 3-Branes,'' 
  Phys.\ Rev.\ D {\bf 54} (1996) 3915 
  [arXiv:hep-th/9602135].\\ 
  I.~R.~Klebanov and A.~A.~Tseytlin, 
``Entropy of Near-Extremal Black p-branes,'' 
  Nucl.\ Phys.\ B {\bf 475} (1996) 164 
  [arXiv:hep-th/9604089]. 
  S.~S.~Gubser, I.~R.~Klebanov and A.~A.~Tseytlin, 
 ``Coupling constant dependence in the thermodynamics of N = 4  supersymmetric 
 Yang-Mills theory,'' 
  Nucl.\ Phys.\ B {\bf 534} (1998) 202 
  [arXiv:hep-th/9805156]. 
 
 
\bibitem{wittenhp} 
  E.~Witten, 
``Anti-de Sitter space, thermal phase transition, and confinement in  gauge 
theories,'' 
  Adv.\ Theor.\ Math.\ Phys.\  {\bf 2} (1998) 505 
  [arXiv:hep-th/9803131]. 
 
 
 
\bibitem{kw} 
  I.~R.~Klebanov and E.~Witten, 
 ``Superconformal field theory on threebranes at a Calabi-Yau  singularity,'' 
  Nucl.\ Phys.\ B {\bf 536} (1998) 199 
  [arXiv:hep-th/9807080]. 
 
\bibitem{gk} 
  S.~S.~Gubser and I.~R.~Klebanov, 
 ``Baryons and domain walls in an N = 1 superconformal gauge theory,'' 
  Phys.\ Rev.\ D {\bf 58} (1998) 125025 
  [arXiv:hep-th/9808075]. 
 
\bibitem{kn} 
  I.~R.~Klebanov and N.~A.~Nekrasov, 
 ``Gravity duals of fractional branes and logarithmic RG flow,'' 
  Nucl.\ Phys.\ B {\bf 574} (2000) 263 
  [arXiv:hep-th/9911096]. 
 
\bibitem{kt} 
  I.~R.~Klebanov and A.~A.~Tseytlin, 
 ``Gravity duals of supersymmetric SU(N) x SU(N+M) gauge theories,'' 
  Nucl.\ Phys.\ B {\bf 578} (2000) 123 
  [arXiv:hep-th/0002159]. 
 
\bibitem{ks} 
  I.~R.~Klebanov and M.~J.~Strassler, 
 ``Supergravity and a confining gauge theory: Duality cascades and 
 chiSB-resolution of naked singularities,'' 
  JHEP {\bf 0008} (2000) 052 
  [arXiv:hep-th/0007191]. 
 
\bibitem{hko} 
C.~P.~Herzog, I.~R.~Klebanov and P.~Ouyang, 
``D-branes on the conifold and N = 1 gauge / gravity dualities,'' 
arXiv:hep-th/0205100. 
 
 
\bibitem{alex} 
  A.~Buchel, 
 ``Finite temperature resolution of the Klebanov-Tseytlin singularity,'' 
  Nucl.\ Phys.\ B {\bf 600} (2001) 219 
  [arXiv:hep-th/0011146]. 
 
\bibitem{105} 
  A.~Buchel, C.~P.~Herzog, I.~R.~Klebanov, L.~A.~Pando Zayas and A.~A.~Tseytlin, 
  ``Non-extremal gravity duals for fractional D3-branes on the conifold,'' 
  JHEP {\bf 0104} (2001) 033 
  [arXiv:hep-th/0102105]. 
 
\bibitem{172} 
  S.~S.~Gubser, C.~P.~Herzog, I.~R.~Klebanov and A.~A.~Tseytlin, 
  ``Restoration of chiral symmetry: A supergravity perspective,'' 
  JHEP {\bf 0105} (2001) 028 
  [arXiv:hep-th/0102172]. 

\bibitem{aby} 
  O.~Aharony, A.~Buchel and A.~Yarom, 
 ``Holographic renormalization of cascading gauge theories,'' 
  Phys.\ Rev.\ D {\bf 72} (2005) 066003 
  [arXiv:hep-th/0506002]. 
 
\bibitem{PT} 
  L.~A.~Pando Zayas and C.~A.~Terrero-Escalante, 
  JHEP {\bf 0609} (2006) 051 
  [arXiv:hep-th/0605170]. 
 
\bibitem{marolf} 
  D.~Marolf, 
 ``Chern-Simons terms and the three notions of charge,'' 
  arXiv:hep-th/0006117. 
 
 
\bibitem{a2} 
  A.~Buchel, 
 ``Transport properties of cascading gauge theories,'' 
  Phys.\ Rev.\ D {\bf 72} (2005) 106002 
  [arXiv:hep-th/0509083]. 
 
 
\bibitem{resolved} 
  L.~A.~Pando Zayas and A.~A.~Tseytlin, 
``3-branes on resolved conifold,'' 
  JHEP {\bf 0011} (2000) 028 
  [arXiv:hep-th/0010088]. 
 
  
\bibitem{cobitemp} 
  O.~Aharony, J.~Sonnenschein and S.~Yankielowicz, 
``A holographic model of deconfinement and chiral symmetry restoration,'' 
  arXiv:hep-th/0604161. 
 
\bibitem{temp1} 
  D.~Mateos, R.~C.~Myers and R.~M.~Thomson, 
``Holographic phase transitions with fundamental matter,'' 
  arXiv:hep-th/0605046. 
 
\bibitem{temp2} 
  P.~Benincasa and A.~Buchel, 
``Hydrodynamics of Sakai-Sugimoto model in the quenched approximation,'' 
  arXiv:hep-th/0605076. 
 
\bibitem{temp3} 
  A.~Parnachev and D.~A.~Sahakyan, 
``Chiral phase transition from string theory,'' 
  arXiv:hep-th/0604173. 
 
\bibitem{temp4} 
  T.~Albash, V.~Filev, C.~V.~Johnson and A.~Kundu, 
``A topology-changing phase transition and the dynamics of flavour,'' 
  arXiv:hep-th/0605088. 
 


\bibitem{PI}
L. A. Pando Zayas, {\it Confinement/Deconfinement Transition in AdS / CFT } , 
http://www.perimeterinstitute.ca/en/Events/\\
Exotic\_States\_of\_Hot\_and\_Dense\_Matter\_and\_their\_Dual\_Description/View\_Lectures/

\bibitem{kapusta}
J. I. Kapusta and C. Gale, Finite-Temperature Field Theory Principles and Applications, 
Cambridge University Press, 2006.

\bibitem{bellac}
M. Bellac, Thermal Field Theory, Cambridge University Press, 1996.



\bibitem{alexuniversality} 
  A.~Buchel, 
  Phys.\ Lett.\  B {\bf 609} (2005) 392 
  [arXiv:hep-th/0408095]. 
 
\bibitem{alexjim}
  A.~Buchel and J.~T.~Liu,
  Phys.\ Rev.\ Lett.\  {\bf 93} (2004) 090602
  [arXiv:hep-th/0311175].

 
\bibitem{herzogtransition} 
  C.~P.~Herzog, 
  Phys.\ Rev.\ Lett.\  {\bf 98} (2007) 091601 
  [arXiv:hep-th/0608151]. 
 
 
\bibitem{braziltransition} 
  C.~A.~Ballon Bayona, H.~Boschi-Filho, N.~R.~F.~Braga and L.~A.~Pando Zayas, 
  arXiv:0705.1529 [hep-th]. 
 
\bibitem{Cai:2007zw}
  R.~G.~.~Cai and J.~P.~Shock,
  arXiv:0705.3388 [hep-th].

\bibitem{Cai:2007bq}
  R.~G.~Cai and N.~Ohta,
  arXiv:0707.2013 [hep-th].
 
\bibitem{smearing} 
  I.~R.~Klebanov and A.~Murugan, 
  JHEP {\bf 0703} (2007) 042 
  [arXiv:hep-th/0701064]. 
 

\bibitem{seattle}
  C.~P.~Herzog, A.~Karch, P.~Kovtun, C.~Kozcaz and L.~G.~Yaffe,
  JHEP {\bf 0607} (2006) 013
  [arXiv:hep-th/0605158].


\bibitem{gubser}
  S.~S.~Gubser,
  Phys.\ Rev.\  D {\bf 74} (2006) 126005
  [arXiv:hep-th/0605182].




\bibitem{herzog}
  C.~P.~Herzog,
  JHEP {\bf 0609} (2006) 032
  [arXiv:hep-th/0605191].


 
\bibitem{alexjet}
  A.~Buchel,
  Phys.\ Rev.\  D {\bf 74} (2006) 046006
  [arXiv:hep-th/0605178].

\bibitem{elenajet}
  E.~Caceres and A.~Guijosa,
  JHEP {\bf 0612} (2006) 068
  [arXiv:hep-th/0606134].




\bibitem{ehk} 
  C.~P.~Herzog, Q.~J.~Ejaz and I.~R.~Klebanov, 
  JHEP {\bf 0502} (2005) 009 
  [arXiv:hep-th/0412193]. 
 
 
\bibitem{ypqcas}
  B.~A.~Burrington, J.~T.~Liu, M.~Mahato and L.~A.~Pando Zayas, 
  JHEP {\bf 0507} (2005) 019 
  [arXiv:hep-th/0504155]. 
 






\bibitem{backreactedflavor}
  F.~Benini, F.~Canoura, S.~Cremonesi, C.~Nunez and A.~V.~Ramallo,
  arXiv:0706.1238 [hep-th].\\
  F.~Benini, F.~Canoura, S.~Cremonesi, C.~Nunez and A.~V.~Ramallo,
  JHEP {\bf 0702} (2007) 090
  [arXiv:hep-th/0612118].



\bibitem{alexofer}
  O.~Aharony, A.~Buchel and P.~Kerner,
  arXiv:0706.1768 [hep-th].



\end{thebibliography}
\end{document}